\documentstyle[12pt,epsf]{article}
\begin{document}

\newcommand{\pa}{$\pi {-}a_1$}
\newcommand{\mch}{\stackrel{\circ}{m}}
\newcommand{\mpio}{\stackrel{\circ}{m}_\pi}
\newcommand{\Lch}{\left(\frac{\Lambda^2}{\Lambda^2+\mch^2}\right)^2}
\newcommand{\nh}{\stackrel{\circ}{h}_1}
\newcommand{\Io}{\stackrel{\circ}{I}_2}
\newcommand{\Do}{\stackrel{\circ}{\delta}}
\newcommand{\nhh}{\stackrel{\circ}{h}_2}

\thispagestyle{empty}

\vspace {-3truecm}

\begin{center}
    {\large PION OBSERVABLES IN THE EXTENDED NJL MODEL \\
       WITH VECTOR AND AXIAL-VECTOR MESONS\footnote{Supported in part by the
        grants from CERN No PCERN C/FAE/74/91; STRDA /C/CEN/564/92;
        CERN/FIS/116/94 and Russian Foundation of Fundamental
        Research No 94-02-03028.}}
\end{center}

\begin{center}{V\'eronique Bernard\\
        {\small\em Centre de Recherches Nucl\'{e}aires,
        Physique Th\'{e}orique, Bat.40}\\
        {\small\em BP 28, F-67037 Strasbourg
        Cedex 2, France}\\
        {\small\em Laboratoire de Physique Th\'eorique, Institut
         de Physique}\\
        {\small\em 3-5, rue de l'Universit\'e,
         F-67084 Strasbourg Cedex, France}\\

\vspace {0.4truecm}

        Alex H. Blin, Brigitte Hiller\\
        {\small\em Centro de F\'{\i}sica Te\'{o}rica, Departamento de
        F\'{\i}sica}\\
        {\small\em da Universidade de Coimbra, P-3000 Coimbra, Portugal}\\

\vspace {0.4truecm}

        Yuri P. Ivanov, Alexander A. Osipov\\
        {\small\em Joint Institute for Nuclear Research,
        Laboratory of Nuclear Problems}\\
        {\small\em 141980 Dubna, Moscow Region, Russia}\\

\vspace {0.4truecm}

        Ulf-G. Mei\ss ner\\
        {\small\em Universit\"at Bonn, Institut f\"ur Theoretische
         Kernphysik}\\
        {\small\em Nussallee 14-16, D-53115 Bonn, Germany}
       }
\end{center}


\vspace{2truecm}

\begin{abstract}
The momentum-space bosonization method of a Nambu and Jona-Lasinio type
model with vector and axial-vector mesons is applied to $\pi\pi$ scattering.
Unlike the case in earlier published papers, we obtain the $\pi\pi$
scattering amplitude using the linear and nonlinear realizations of
chiral symmetry and fully taking into account the momentum dependence
of meson vertices. We show the full physical equivalence between these
two approaches. The chiral expansion procedure in this model is
discussed in detail. Chiral expansions of the quark mass, pion mass and
constant $f_\pi$ are obtained. The low-energy $\pi \pi $ phase shifts are
compared to the available data. We also study the scalar form factor of
the pion.
\end{abstract}

\vspace{2truecm}

\noindent CRN 95-25

\noindent TK 95 18

\newpage
\section{Introduction}

Coloured quarks and gluons are the main objects in the fundamental theory of
the strong interactions, quantum chromodynamics (QCD), which are not directly
observed in experiment. At intermediate and low energies the laws
governing interactions of these fundamental particles bring about complex
rearrangements inside a quark-gluon system which eventually result in
two-quark and three-quark colourless formations---hadrons. A second transition
appears at a comparable scale,
 namely the spontaneous chiral symmetry breakdown,
leading to the pseudo-Goldstone boson octet. These phenomena cannot be
consistently described within perturbation theory in the strong coupling
constant  $\alpha_s$ which is no longer small at the typical hadronic energy
scale. Therefore, it is mandatory to develop nonperturbative methods to solve
these problems.

There are few nonperturbative methods employed in particle physics.
Noteworthy are lattice field theory,
quantization on the light cone, and QCD sum rules. Despite all their
advantages, these approaches have not yet lead to a unified description of
confinement
and spontaneous chiral symmetry breakdown $(\chi SB)$.
Both phenomena are supposed to
arise from colour gauge interactions. Yet, the complex process of hadron
formation remains obscure. This stimulates numerous attempts to construct
models aimed at describing individual major features of the process.

One can find explanation of spontaneous symmetry breakdown  in the
Nambu--Jona-Lasinio model, chiral quark model, approaches with stochastic
vacuum fields, dual QCD, etc. A large class of such models involve path
integral
methods for calculation of Green functions that approximate QCD in the
infrared region. The guiding principle is iterative integration first with
respect to rapidly varying fields and then with respect to slowly varying
ones, an idea borrowed from the general theory of phase transitions.
According to this idea, slow modes  mainly determine the
asymptotic behaviour of the system at small energies and momenta (infrared
asymptotics) and require thorough study in terms of collective
variables. The contribution from ``fast'' fields is less substantial and
can be taken into account by perturbation theory. The
Nambu--Jona-Lasinio model (NJL) \cite{1}, on which we concentrate below, is
one of these schemes. The boundary between ``slow'' and ``fast'' fields is
largely conventional being dictated by physical considerations. The chiral
$SU(3)_L\otimes SU(3)_R$ symmetry of the QCD lagrangian is broken at
energies of the order of $\Lambda_{\chi SB}\sim 1\,$GeV if small quark
masses are ignored. This is what allows one to relate $\Lambda_{\chi SB}$
to the boundary between ``large'' and ``small'' momenta when looking into
infrared QCD asymptotics from the symmetry point of view.

Chiral symmetry breakdown is a fundamental issue because it is related
to the problem of elementary particle masses and chiral symmetry  lends
itself  to  a controlable expansion (in energy) at
energies below $\Lambda_{\chi SB}$.  At these energies
the QCD vacuum
becomes highly non-perturbative due to the condensation of quark-antiquark
pairs as a consequence of the
spontaneous  symmetry violation. It is believed that any serious model
of the strong interactions has to account for these facts. This is the
case for the celebrated NJL model which was originally formulated not
in terms of quark but rather  nucleonic degrees of freedom. In what follows
we will be concerned with a modern extension based on the quark description.

Assuming that the NJL model in the form that involves all essential QCD
symmetries may be a reasonable approximation of QCD in the intermediate
energy range, we develop a new theoretical method. It allows arbitrary
N-point functions describing strong and electroweak interactions of basic
meson states to be calculated through the effective four-quark interaction of
the NJL model \cite{2}. Now there are two approaches to the problem. One
is based on the classical treatment of four-fermion Lagrangians (the
Hartree--Fock approximation plus Bethe -- Salpeter type equations for bound
states) \cite{3}-\cite{7}. The other involves explicit separation of
collective degrees of freedom in the functional integral of the theory
(bosonization) \cite{8}-\cite{14}. In this framework one usually derives
the Lagrangian based on a derivative expansion
instead of directly calculating the Green functions.
For technical reasons, such a derivative expansion can only be performed
to the first few orders.
 {\it We have constructed such a formal scheme that at the same
time incorporates bosonization and has all advantages of the pure fermionic
approach.} It allows to combine the advantages of both of the abovementioned
methods. The results of our more general method can be directly compared
with the  ones based on standard bosonization by performing a chiral expansion
based on the nonlinear Bethe -- Salpeter equations.

Originally formulated in terms of boson variables linearly transforming
with respect to the chiral group action, the method can be extended to
nonlinear realization of symmetry. This is one of the significant advantages
of the approach as compared with Hartree and Fock's classical technique,
which is linear is this sense. Now chiral symmetry consequences are often
investigated via nonlinear chiral Lagrangians. They are a series in energy
$(E)$, which is equivalent to normal expansion with an increasing number
of derivatives. As a rule, investigations are mostly  confined to the
${\cal O}(E^4)$ approximation. Recently a program like this was used for
the extended NJL model \cite{14}. In the present paper we show how
one can generalize these results without any reference to an expansion
in $E^2$.

Apart from pure theoretical developments to generalize the results of
\cite{14}, we carry out explicit calculations. Our primary concern
is the $\pi\pi$ scattering amplitude. We shall calculate it both in the
linear and nonlinear approaches using the standard Lagrangian of the
extended NJL model with scalar, pseudoscalar, vector, and axial-vector
four-quark interactions. We show that there is no difference between these
two approaches. The expressions for the total amplitude $A(s,t,u)$ are
the same in both cases. It has to be so because the same Lagrangian is
used. We consider this point in detail to show the
self-consistency of our method.

The paper is organized as follows. In the next section we introduce the
Lagrangian and notation as well as review the main steps of
the momentum-space bosonization technique.
We then consider the nonlinear realization of
chiral symmetry. We shall show that in this case equations for masses of
bound states coincide with analogous equations in the linear approach. The
expression for the weak pion decay constant $f_\pi$ does not change
either. All this allows us to get common rules for constructing chiral
expansions  in the linear and nonlinear approaches.
In section 3 we derive the $\pi\pi$ scattering amplitudes and the main
low-energy characteristics of the process in question. Theoretical
derivations of the previous section will be the starting point. At first
we shall use the model with linear realization of chiral symmetry. In this
section our calculations are an extension of earlier investigations [2,
15]. The corresponding amplitude includes the exchange of scalar and
vector mesons as well as quark box diagrams. We give explicit expressions
for this amplitude in terms of a few standard integrals. We obtain the
low-energy expansion for the scattering amplitude and recover Weinberg's
\cite{16} formula (to lowest order in
the energy expansion). Examining the model arising from classification
of collective excitations based on nonlinear representations of the chiral
group, we arrive at the same expression for the $\pi\pi$ scattering
amplitude. We demonstrate this point by direct calculations and comparing
the different contributions. In this part of the paper we not only generalize
the known result of \cite{14}, we also prove the equivalence of this approach
with the other known results. Our method looks like a bridge between
different
NJL models. Finally we present our results, comparing different variants of
the NJL model with experiment and chiral perturbation results.
In section 4 we calculate the scalar form factor of the pion.
We pay particular
attention to its relation to the value of the constituent quark mass.
We conclude with a summary of our results.

\section{The theoretical background}
In Ref.\cite{17} the extended NJL model  was investigated and it
was shown how to calculate
Fourier transforms of $N$-point Green functions by a simple
method which was proposed in \cite{2}. This calculation
was done in the linear realization of
chiral symmetry. In this section we are developing the method to apply it
to the extended NJL model with nonlinear realization of chiral symmetry.
To describe collective excitations we shall use induced representations of
the chiral group. Calculations of quark loop diagrams will involve
renormalization  of collective variables,
as is typical of our approach. As a result, we
are able to  calculate amplitudes of physical processes to
any accuracy in $E^2$ in models with nonlinear realization of chiral
symmetry.

\subsection{The linear approach}
In this section we summarize previous results (see also paper \cite{17}).
The reader who is familiar with the momentum-space bosonization method can
use this section as a collection of our notation and main formulae to be
found in this paper. The main idea of this method consists in the
construction of special bosonic variables to be used for the description
of the observable mesonic states. As a result, it extends the usual
treatment of bosonized NJL models, which was formulated in \cite{8}
and developed in [9,10,12]. The standard approach is essentially
linked to the derivative expansion of the effective meson Lagrangian.
The momentum-space bosonization method does not use this approximation.
It involves the bosonization procedure and has all the advantages of the pure
fermionic approach (Hartree-Fock plus Bethe-Salpeter approximation).
Instead of step-by-step construction of the effective Lagrangian
describing the dynamics of collective excitations we develop a method that
allows direct calculation of amplitudes for particular physical processes.
The amplitudes derived accumulate the entire information on the process under
investigation, just as if we had a total effective meson Lagrangian of the
bosonized  NJL model (in the one-loop approximation).

Consider the extended $SU(2)_L\otimes SU(2)_R$ NJL Lagrangian with the
local four-quark interactions
\begin{eqnarray}
\label{la:1}
L(q)=\bar{q}(i\gamma^\mu\partial_\mu -\widehat{m})q
    \!&+&\!\frac{G_S}{2}
       \left[(\bar{q}\tau_aq)^2+(\bar{q}i\gamma_5\tau_aq)^2
       \right]\nonumber \\
    \!&-&\!\frac{G_V}{2}
       \left[(\bar{q}\gamma^{\mu}\tau_aq)^2+
             (\bar{q}\gamma_{5}\gamma^{\mu}\tau_aq)^2
       \right],
\end{eqnarray}
where $\bar{q}=(\bar{u}, \bar{d})$ are coloured $(N_c=3)$ current quark
fields with current mass $\widehat{m}=\mbox{diag}(\widehat{m}_u,
\widehat{m}_d)$, $\tau_a=(\tau_0, \tau_i),\, \tau_0=I,\, \tau_i\, (i=1, 2,
3)$ are the Pauli matrices of the flavour group $SU(2)_f$. The constants of
the four-quark interactions are $G_S$ for the scalar and pseudoscalar cases,
$G_V$ for the vector and the axial-vector cases. The current mass term
explicitly breaks the $SU(2)_L\otimes SU(2)_R$ chiral symmetry of the
Lagrangian (\ref{la:1}). In what follows, we shall only consider the
isospin symmetrical case $\widehat{m}_u= \widehat{m}_d=\widehat{m}$. With
boson fields introduced in the standard way, the Lagrangian takes the form
\begin{eqnarray}
\label{la:2}
L(q, \bar{\sigma}, \tilde{\pi}, \tilde{v}, \tilde{a})\!
    &=&\!\bar{q}\left(i\gamma^\mu\partial_\mu -\widehat{m}
             +\bar{\sigma}
             +i\gamma_5\tilde{\pi}
             +\gamma^\mu\tilde{v}_\mu
             +\gamma_5\gamma^\mu\tilde{a}_\mu
             \right)q\nonumber \\
    \!&-&\!\frac{\bar{\sigma}^2_a+\tilde{\pi}^2_a}{2G_S}
      +\frac{\tilde{v}^2_{\mu a}+\tilde{a}^2_{\mu a}}{2G_V}.
\end{eqnarray}
Here $\bar{\sigma}=\bar{\sigma}_a\tau_a,\ \tilde{\pi}=\tilde{\pi}_a\tau_a,
\ \tilde{v}_{\mu}=\tilde{v}_{\mu a}\tau_a,\ \tilde{a}_{\mu}=\tilde{a}_{\mu a}
\tau_a$. The vacuum expectation value of the scalar field $\bar{\sigma}_0$
turns out to be different from zero $(<\bar{\sigma}_0>\neq 0)$. To obtain
the physical field $\tilde{\sigma}_0$ with $<\tilde{\sigma}_0>=0$ one
performs a field shift leading to a new quark mass $m$ to be identified with
the mass of the constituent quarks
\begin{equation}
\label{la:3}
\bar{\sigma}_0-\widehat{m}=\tilde{\sigma}_0-m,\qquad
\bar{\sigma}_i=\tilde{\sigma}_i,
\end{equation}
where $m$ is determined from the gap equation (see (\ref{la:9}) below).

Let us integrate out the quark fields in the generating functional
associated with the Lagrangian (\ref{la:2}). Evaluating the resulting quark
determinant by a loop expansion one obtains
\begin{eqnarray}
\label{la:4}
L(\tilde{\sigma}, \tilde{\pi}, \tilde{v}, \tilde{a})\!
    &=\!&-i\mbox{Tr}\ln
      \left[1+
      (i\gamma^\mu\partial_\mu -m)^{-1}
      (\tilde{\sigma}+i\gamma_5\tilde{\pi}+\gamma^\mu\tilde{v}_\mu
      +\gamma_5\gamma^\mu\tilde{a}_\mu )
      \right]_{\Lambda}\nonumber \\
  \!&-&\!\frac{\bar{\sigma}^2_a+\tilde{\pi}^2_a}{2G_S}
      +\frac{\tilde{v}^2_{\mu a}+\tilde{a}^2_{\mu a}}{2G_V}.
\end{eqnarray}
The NJL model belongs to the set of nonrenormalizable theories. Hence, to
define it completely as an effective model, a regularization scheme must
be specified to deal with the quark-loop integrals in harmony with general
symmetry requirements. As a result, an  additional parameter $\Lambda$
appears, which characterizes the scale of the quark-antiquark forces
responsible for the dynamic chiral symmetry breaking. From the meson mass
spectrum it is known that $\Lambda\sim 1\,\hbox{GeV}$. Here, we shall make
use of the a modified
Pauli--Villars \cite{18} regularization, which preserves gauge
invariance and chiral symmetry.
In this form it was used in \cite{19}-\cite{20}. The
Pauli--Villars cut-off $\Lambda$ is introduced in the following way,
\begin{equation}
\label{la:5}
         e^{-im^2z}
         \rightarrow R(z)=e^{-im^2z}\left[1-(1+iz\Lambda^2)
         e^{-iz\Lambda^2}\right],
\end{equation}
\begin{equation}
\label{la:6}
         m^2e^{-im^2z}
         \rightarrow iR'(z)
         =m^2R(z)-iz\Lambda^4e^{-iz(\Lambda^2+m^2)},
\end{equation}
where only one Pauli--Villars regulator has been introduced. In this case
the expressions for the basic loop integrals $I_i$ coincide with those
obtained by the usual covariant cut-off scheme. Let us give our
definitions for these integrals. To simplify the formulae we introduce the
following notation
\begin{equation}
\label{la:7}
\Delta (p)=\frac{1}{p^2-m^2},\qquad
\tilde{d}^4q=\frac{d^4q}{(2\pi )^4}.
\end{equation}
Then we have
\begin{eqnarray}
\label{la:8}
&&I_1=iN_c\int\tilde{d}^4q\Delta (q)
    =\frac{N_c}{(4\pi )^2}\left[\Lambda^2-
    m^2\ln\left(1+\frac{\Lambda^2}{m^2}\right)\right],\\
&&I_2(p^2)=-iN_c\int\tilde{d}^4q\Delta (q)\Delta (q+p) \nonumber \\
&&\ \ \ \ \ \ \ \ \
          =\frac{N_c}{16\pi^2}\int_0^1dy\int_0^{\infty}\frac{dz}{z}R(z)
                e^{\frac{i}{4}zp^2(1-y^2)}, \\
&&J_2(p^2)=\frac{3N_c}{32\pi^2}\int_0^1dy(1-y^2)\int_0^{\infty}\frac{dz}{z}
                R(z)e^{\frac{i}{4}zp^2(1-y^2)},\\
&&I_3(p_1, p_2)=-iN_c\int\tilde{d}^4q\Delta (q)\Delta (q+p_1)\Delta
                                                           (q+p_2),\\
&&I_4(p_1,p_2,p_3)=-iN_c\int\tilde{d}^4q\Delta (q)\Delta (q+p_1)
                                        \Delta (q+p_2)\Delta (q+p_3).
\end{eqnarray}

Consider the first terms of the logarithm expansion in (\ref{la:4}). From
the requirement for the terms linear in $\tilde{\sigma}$ to vanish we get
a modified gap equation
\begin{equation}
\label{la:9}
m-\widehat{m}=8mG_SI_1.
\end{equation}

The terms quadratic in the boson fields lead to the amplitudes
\begin{eqnarray}
\label{la:10}
\Pi^{PP}(p^2)
    &=&\left[8I_1-G^{-1}_S+p^2g^{-2}(p^2)\right]
       \varphi^+_P\varphi^-_P,\\
\Pi^{SS}(p^2)
    &=&\left[8I_1-G^{-1}_S+(p^2-4m^2)g^{-2}(p^2)\right]
       \varphi^+_S\varphi^-_S,\\
\Pi^{VV}(p^2)
    &=&\left[g^{\mu\nu}G_V^{-1}+4(p^{\mu}p^{\nu}-g^{\mu\nu}p^2)
       g_V^{-2}(p^2)\right]\varepsilon^{*V}_{\mu}(p)
       \varepsilon^V_{\nu}(p),\\
\Pi^{AA}(p^2)
    &=&\left[g^{\mu\nu}\left( G_V^{-1}+4m^2g^{-2}(p^2)\right)
       \right.\nonumber \\
    & &\quad +\left. 4(p^{\mu}p^{\nu}-g^{\mu\nu}p^2)
       g_V^{-2}(p^2)\right]\varepsilon^{*A}_{\mu}(p)
       \varepsilon^A_{\nu}(p),\\
\Pi^{PA}(p^2)
    &=&2img^{-2}(p^2)p^{\mu}\varepsilon^{*A}_{\mu}(p)
    \varphi^-_P,\\
\Pi^{AP}(p^2)
    &=&-2img^{-2}(p^2)p^{\mu}\varepsilon^A_{\mu}(p)
    \varphi^+_P.
\end{eqnarray}
Here $\varepsilon^V_{\mu}(p), \varepsilon^A_{\mu}(p)$
are the polarization vectors of the vector and axial-vector fields.
We have introduced the symbols $\varphi^-_P=1$ and $\varphi^-_S=1$ to
explicitly show the pseudoscalar and scalar field contents of the
pertinent two-point functions. The functions $g(p^2)$ and $g_V(p^2)$
are determined by the following integrals
\begin{equation}
\label{la:11}
g^{-2}(p^2)= 4I_2 (p^2),
\end{equation}
\begin{equation}
\label{la:12}
g^{-2}_V(p^2)=\frac{2}{3}J_2 (p^2).
\end{equation}
Let us diagonalize the quadratic form $(14)+(17)+(18)+(19)$
by redefining the axial fields
\begin{eqnarray}
\label{la:13}
      \varepsilon^{A}_{\mu}(p)
      &\rightarrow &
      \varepsilon^{A}_{\mu}(p)-i\beta (p^2)p_{\mu}
      \varphi^{-}_{P}, \\
      \varepsilon^{*A}_{\mu}(p)
      &\rightarrow &
      \varepsilon^{*A}_{\mu}(p)+i\beta (p^2)p_{\mu}
      \varphi^{+}_{P}.
\end{eqnarray}
This determines the function $\beta (p^2)$,
\begin{equation}
\label{la:14}
\beta (p^2)=\frac{8mI_2(p^2)}{G_V^{-1}+16m^2I_2(p^2)}.
\end{equation}
Consequently, one has no more mixing
between pseudoscalar and axial-vector fields.
The self-energy of the pseudoscalar field takes the form
\begin{equation}
\label{la:15}
\Pi^{PP}(p^2)=
    \left[8I_1-G^{-1}_S+p^2g^{-2}(p^2)
    \left(1-2m\beta(p^2)\right)\right]
    \varphi^+_P\varphi^-_P.
\end{equation}
Now we can construct special boson variables that will describe
the observed mesons. These field functions $\phi$\footnote{Here and in the
following we will use the common symbol $\phi$ for the all set of
meson fields: $\pi _a, \sigma _a, v_a, a_a$.}
correspond to bound quark -- antiquark states and are derived via the
following transformations
\begin{eqnarray}
\label{la:16}
\tilde{\pi}^a(p)&=&Z^{-1/2}_{\pi}g_{\pi}(p^2)\pi^a(p),\\
\tilde{\sigma}^a(p)&=&Z^{-1/2}_{\sigma}g(p^2)\sigma^a(p),\\
\tilde{v}^a(p)&=&\frac{1}{2}Z^{-1/2}_{v}g_V(p^2)v^a(p),\\
\tilde{a}^a(p)&=&\frac{1}{2}Z^{-1/2}_{a}g_V(p^2)a^a(p),
\end{eqnarray}
where
\begin{equation}
\label{la:17}
g_{\pi}(p^2)=\frac{g(p^2)}{\sqrt{1-2m\beta (p^2)}}
            =g(p^2)\sqrt{1+16m^2G_VI_2(p^2)}.
\end{equation}
The new bosonic fields have the self-energies
\begin{eqnarray}
\label{la:18}
\Pi ^{\pi , \sigma}_{ab}(p^2)\!&=&\!\delta _{ab}Z^{-1}_{\pi , \sigma}
    \left[p^2-m^2_{\pi , \sigma}(p^2)\right], \nonumber \\
\Pi ^{v, a}_{\mu\nu , ab}(p^2)\!&=&\!\delta _{ab}Z^{-1}_{v, a}
    \left\{p_{\mu}p_{\nu}-g_{\mu\nu}\left[ p^2-m^2_{v, a}(p^2)\right]
    \right\}.
\end{eqnarray}
The $p^2$-dependent masses are equal to
\begin{eqnarray}
\label{la:19}
m^2_{\pi}(p^2)\!&=&\! (G^{-1}_S-8I_1)g^2_{\pi}(p^2)
               =\widehat{m}(mG_S)^{-1}g^2_\pi (p^2),\\
m^2_{\sigma}(p^2)\!
              &=&\!\left[1-2m\beta (p^2)\right]m^2_{\pi}(p^2)+4m^2,\\
m^2_v(p^2)\!&=&\!\frac{g^2_V(p^2)}{4G_V}=\frac{3}{8G_VJ_2(p^2)},\\
m^2_a(p^2)\!&=&\! m^2_v(p^2)+6m^2\frac{I_2(p^2)}{J_2(p^2)}.
\end{eqnarray}
These equations coincide with the conditions for appearance of
quark -- antiquark bound state as deduced in the pure fermion approach from
analysis of the Bethe -- Salpeter equations.

The constants $Z_\phi$ are determined by the requirement
that the inverse meson field propagators $\Pi^\phi$
satisfy the normalization conditions
\begin{eqnarray}
\label{la:20}
\Pi ^{\pi , \sigma}(p^2)\!&=&\!
     p^2-m^2_{\pi , \sigma}+{\cal O}\left(
    (p^2-m^2_{\pi , \sigma})^2\right), \nonumber \\
\Pi ^{v, a}_{\mu\nu}(p^2)\!&=&\! -g_{\mu\nu}\left[
     p^2-m^2_{v, a}+{\cal O}\left(
    (p^2-m^2_{v, a})^2\right)\right],
\end{eqnarray}
around the physical mass points $p^2=m^2_\phi$,
respectively. The conditions (\ref{la:20}) lead to the values
\begin{eqnarray}
\label{la:21}
Z_{\pi}\!&=&\! 1+\frac{m^2_{\pi}[1-2m\beta (m^2_\pi )]}{I_2(m^2_\pi )}
   \frac{\partial I_2(p^2)}{\partial p^2}
   \bigg\vert_{p^2=m^2_{\pi}},\\
Z_{\sigma}\!&=&\! 1+\frac{
   m^2_{\sigma}-4m^2}{I_2(m^2_{\sigma})}
   \frac{\partial I_2(p^2)}{\partial p^2} \bigg\vert_{p^2=
   m^2_{\sigma}},\\
Z_v\!&=&\! 1+\frac{m^2_v}{J_2(m^2_v)}
   \frac{\partial J_2(p^2)}{\partial p^2} \bigg\vert_{p^2=
   m^2_v},\\
Z_a\!&=&\! 1+\frac{m^2_a}{J_2(m^2_a)}
   \frac{\partial J_2(p^2)}{\partial p^2} \bigg\vert_{p^2=
   m^2_a}-\frac{6m^2}{J_2(m^2_a)}
   \frac{\partial I_2(p^2)}{\partial p^2} \bigg\vert_{p^2=
   m^2_a}.
\end{eqnarray}
In the following, when omitting an argument of a running coupling constant
or a running mass, we always assume that its value is taken on the
mass-shell of the corresponding particle. The symbol of this particle
will be used for that. For example, on the pion mass-shell
$m^2_\pi (p^2\! =\! m^2_\pi )=m^2_\pi ,\ \beta(m^2_\pi )=\beta_\pi$ and
so on.

Using the expressions (\ref{la:18}), one can obtain the two-point meson
Green functions $\Delta^\phi (p)$. For example, in the scalar
and vector field case the relations
\begin{equation}
\label{la:22}
\Pi^{\sigma}_{ab}(p^2)\Delta^{\sigma}_{bc}(p^2)=\delta_{ac}, \quad
\Pi^v_{\mu\nu ,ab}(p^2)\Delta^{v, \nu\sigma}_{bc}(p)=\delta_{ac}
      \delta^{\sigma}_\mu
\end{equation}
give
\begin{equation}
\label{la:23}
\Delta^{\sigma}_{ab}(p^2)=\frac{\delta_{ab}Z_\sigma}{p^2-m^2_\sigma (p^2)},
\quad
\Delta^{v, \mu\nu}_{ab}(p)=\frac{\delta_{ab}Z_v}{m^2_v(p^2)}
\frac{p^\mu p^\nu -g^{\mu\nu}m^2_v(p^2)}{p^2-m^2_v(p^2)}.
\end{equation}
This picture corresponds to the calculations in the framework of the
pure fermionic NJL model where the Bethe--Salpeter equation sums an
infinite class of fermion bubble diagrams.

\subsection{The non-linear approach}
Phenomenological meson fields with appropriate transformation properties
under a nonlinear action of the chiral
group can be introduced as follows \cite{wein}, \cite{cwz}. Let $G$ be a
continuous symmetry group of the initial Lagrangian and $H$ a maximum
subgroup of group $G$ which leaves the vacuum invariant. Then an arbitrary
transformation of the group $G$ can be represented as $G=K(\zeta )H(\eta )$,
where $\zeta , \eta$ are the parameters determining the parametrization of
the $G$ group space. Acting from the left on the $G$ group element by an
arbitrary transformation of the same group
$G(g)K(\zeta )H(\eta )=K(\zeta ')H(\eta ')$,
one can find out how the parameters $\zeta$ and $\eta$ are transformed
under transformations of the group. It is essential that in this case the
transformation for $\zeta$ does not involve the parameters $\eta$:
$\zeta '=\zeta '(\zeta , g)$. Each parameter $\zeta^i$ is associated with
a local Goldstone field $\pi^i(x)$ so that the local fields $\pi^i(x)$
obey the transformation rule
\begin{equation}
\label{tb:1}
\pi^{i'}(x)=\zeta^{i'}(\pi^i(x), g).
\end{equation}
The non-linear transformation of the group $G$ on the matter fields is
constructed in the following way,
\begin{equation}
Q \rightarrow Q' = h(\pi ,g) Q ,
\end{equation}
\begin{equation}
R \rightarrow R' = h(\pi ,g) R h^\dagger (\pi ,g) ,
\end{equation}
where $Q$, are the quark fields, $R$, the vector, axial-vector or scalar H
multiplets, and $h(\pi ,g) \!\in\!  H$.

Let us consider the Lagrangian (\ref{la:2}). In this case $G=SU(2)_L\otimes
SU(2)_R$. The quark field $q(x)$ can be represented as
$q(x)=q_L(x)+q_R(x)$, where $q_L(x)=P_Lq(x),\ q_R(x)=P_Rq(x)$. The
projection operators $P_{L,R}$ are $P_{R,L}=(1\pm\gamma_5)/2$. The fields
$q_{L,R}(x)$ transform linearly under action of chiral subgroups
$SU(2)_{L,R}$:
\begin{equation}
\label{tb:2}
q_{L}(x)\rightarrow g_{L}(x)q_{L}(x), \qquad
 q_{R}(x)\rightarrow g_{R}(x)q_{R}(x).
\end{equation}
Let us introduce the notation $\bar{\sigma}_a+i\gamma_5\tilde{\pi}_a=M_a$.
Then we can write $\bar{\sigma}+i\gamma_5\tilde{\pi}=MP_R+M^\dagger P_L$.
Now let us represent the complex $2\times 2$ matrix $M=M_a\tau_a$ as a
product of the unitary matrix $\xi$ and the Hermitian matrix $S$
\begin{equation}
\label{tb:3}
M=\xi S\xi .
\end{equation}
The matrix $\xi$ is parametrized by Goldstone fields, and its
transformation law corresponds to the nonlinear transformation
(\ref{tb:1})
\begin{equation}
\label{tb:4}
\xi\rightarrow g_L(x)\xi h^\dagger (\pi , g_{L,R})=
    h(\pi , g_{L,R})\xi g^\dagger_R(x).
\end{equation}
The map $\xi (\pi )\! : G/H\rightarrow G$ is thus the local section of the
principal $H$-bundle $G\rightarrow G/H$, where $h(\pi , g)\!\in\! H$ is a
compensating $H$ transformation which brings us back to our canonical
choice for coset representative in the new coset specified by $\pi '$.
The exponential parametrization $\xi (\pi )=\exp [i\pi /(2F)]$ corresponds to
the choice of a normal coordinate system in the coset space $G/H$.
New quark variables
\begin{equation}
\label{tb:5}
Q_R=\xi q_R,\qquad Q_L=\xi^\dagger q_L,\qquad Q=Q_R+Q_L
\end{equation}
are transformed by the nonlinear representation of the group
$G$, eq.(44),  and can be used to describe the
constituent quark fields in the approach under consideration.
Let us rewrite the Lagrangian (\ref{la:2}) as
\begin{eqnarray}
\label{tb:6}
L&=&{\bar Q}\left[i\gamma^\mu\nabla_\mu +S-\frac{1}{2}(\Sigma
             +\gamma_5\Delta )+\gamma^\mu\left(W_\mu^{(+)}
             -\gamma_5W_\mu^{(-)}\right)\right]Q
             \nonumber \\
 &-&\frac{1}{4G_S}\mbox{\rm Tr}S^2+\frac{1}{4G_V}\mbox{\rm Tr}
    \left(W_\mu^{(+)}W_\mu^{(+)}+W_\mu^{(-)}W_\mu^{(-)}\right).
\end{eqnarray}
Here we use the following notations:
\begin{eqnarray}
\label{tb:7}
&&\Sigma =\xi^\dagger\widehat{m}\xi^\dagger +\xi\widehat{m}\xi , \qquad
  \Delta =\xi^\dagger\widehat{m}\xi^\dagger -\xi\widehat{m}\xi , \\
&&\nabla_\mu =\partial_\mu +\Gamma_\mu -\frac{i}{2}\gamma_5\xi_\mu , \\
&&\Gamma_\mu =\frac{1}{2}\left(\xi\partial_\mu\xi^\dagger +
                              \xi^\dagger\partial_\mu\xi\right), \quad
  \xi_\mu =i(\xi\partial_\mu\xi^\dagger -
             \xi^\dagger\partial_\mu\xi ).
\end{eqnarray}
To describe the vector $W_\mu^{(+)}$ and axial-vector $W_\mu^{(-)}$ mesons we
use new variables
\begin{equation}
\label{tb:8}
W_\mu^{(\pm )}=\frac{1}{2}\left[\xi^\dagger
               \left(\tilde{v}_\mu +\tilde{a}_\mu\right)\xi\pm
            \xi\left(\tilde{v}_\mu -\tilde{a}_\mu\right)\xi^\dagger\right].
\end{equation}
We take the matrix $S$ in the form $S=\widehat{m}-m+s(x)$. Then the
condition for the tadpole not to appear (after integration over quark
variables) in the case of the scalar field $s(x)$ will be the familiar gap
equation (\ref{la:9}).

Among the terms quadratic in meson fields only the amplitudes
\begin{eqnarray}
\label{tb:9}
\Pi^{PP}(p^2)&=&\frac{4}{F^2}\left[2\widehat{m}(\widehat{m}-m)I_1+
               (m-\widehat{m})^2p^2I_2(p^2)\right]\varphi^i\varphi^j, \\
\Pi^{PA}(p^2)&=&-\frac{8im}{F}(m-\widehat{m})p^\mu\epsilon^{j*}_\mu
               (p)I_2(p^2)\varphi^i, \\
\Pi^{AP}(p^2)&=&\frac{8im}{F}(m-\widehat{m})p^\mu\epsilon^{i}_\mu
               (p)I_2(p^2)\varphi^j
\end{eqnarray}
will be different than in the linear case (see (\ref{la:10})-(19)).
Let us consider a standard replacement in a case like this
\begin{equation}
\label{tb:10}
\epsilon^i_\mu (p)\rightarrow\epsilon^i_\mu (p)+
               ip_\mu\tilde{\beta}\varphi^i, \quad
\epsilon^{j*}_\mu (p)\rightarrow\epsilon^{j*}_\mu (p)-
               ip_\mu\tilde{\beta}\varphi^j.
\end{equation}
The condition for cancellation of nondiagonal terms (absence of the
pseudoscalar -- axial-vector transition) fixes the form of the function
$\tilde{\beta}(p^2)$:
\begin{equation}
\label{tb:11}
\tilde{\beta}(p^2)=\frac{8m(m-\widehat{m})I_2(p^2)}{F[G^{-1}_V
                  +16m^2I_2(p^2)]}.
\end{equation}
The self-energy of the pseudoscalar mode takes the form
\begin{equation}
\label{tb:12}
\Pi^{PP}(p^2)=\frac{4(m-\widehat{m})^2}{F^2}\left[
              \frac{2\widehat{m}}{\widehat{m}-m}I_1+p^2I_2(p^2)
              \left(1-\frac{2m\tilde{\beta}(p^2)F}{m-\widehat{m}}\right)
              \right]\varphi^i\varphi^j.
\end{equation}
Thus, in the nonlinear case only the pseudoscalar field $\pi (p)$ will have a
renormalization other than in the linear approach. The physical field
$\pi^{ph}(p)$ should be introduced by the following replacement
\begin{equation}
\label{tb:13}
\pi (p)=Z^{-1/2}_\pi\tilde{g}_\pi (p^2)\pi^{ph}(p).
\end{equation}
For other fields the transformations remain unchanged (see (27)-(29)). The
function $\tilde{g}_\pi (p^2)$ is determined from (\ref{tb:12})
\begin{equation}
\label{tb:14}
\tilde{g}^2_\pi (p^2)=\frac{F^2[1+16m^2G_VI_2(p^2)]}
                           {4(m-\widehat{m})^2I_2(p^2)}
                     =\left(\frac{Fg_\pi}{m-\widehat{m}}\right)^2.
\end{equation}
Hence it follows in particular that all equations for the masses of collective
modes coincide with the ones in the linear case (\ref{la:19})-(35).
This also applies to the equation of the pion mass, as seen from the chain of
transformations
\begin{equation}
\label{tb:15}
m^2_\pi (p^2)=\frac{8\widehat{m}}{F^2}(m-\widehat{m})\tilde{g}^2_\pi (p^2)I_1
             =\frac{\widehat{m}[1+16m^2G_VI_2(p^2)]}{4mG_SI_2(p^2)},
\end{equation}
where we employed the gap equation (\ref{la:9}). The form factor $f_\pi
(p^2)$ appearing in the vertex of the weak pion decay
$\pi\rightarrow\ell\nu_\ell$ is easily found to be
\begin{equation}
\label{tb:16}
f_\pi (p^2)=\frac{4Z^{-1/2}_\pi m(m-\widehat{m})\tilde{g}_\pi (p^2)I_2(p^2)}
                 {F[1+16m^2G_VI_2(p^2)]}=\frac{m}{\sqrt{Z_\pi}g_\pi (p^2)}.
\end{equation}
The latter equality reveals that it coincides with a similar expression
derived in the linear model \cite{17}. Since the Bethe--Salpeter equation
for the masses of collective states and the expression for the constant $f_\pi$
coincide in the two approaches under consideration, there exists a unified
approach to construct  chiral expansions.

\subsection{Chiral expansion}
At low energies the behaviour of scattering amplitudes or matrix elements
for currents can be described in terms of Taylor expansions in powers of
momenta. Yet, singularities, arising from the presence of light pseudoscalar
particles in the theory, restrict the applicability of  Taylor expansions.
One has to take into account all these singularities to extend the
divergence region for the series in momenta. This is possible because it
is quite clear why the pion mass is small. The pion is a Goldstone boson
and its mass is expressed in terms of current quarks which make it
different from zero. Current quark masses are small and can be taken into
account through perturbation theory. New combined Taylor expansion in
momenta and current quark masses arising in this case form the basis of
chiral perturbation theory \cite{21}.

Let $\mch$ and $f$ be the values of the constituent quark mass $m$ and the
pion decay constant $f_\pi$ in the chiral limit where $\widehat{m}=0$. In
this case the pion mass is zero. The chiral series for this quantity
begins with a term linear in $\widehat{m}$:
\begin{equation}
\label{ce:1}
\stackrel{\circ}{m}^2_\pi=\frac{\widehat{m}\mch}{G_Sf^2}.
\end{equation}
The weak pion decay constant $f_\pi$ is (see eq.(64))
\begin{equation}
\label{ce:2}
f_\pi =\frac{m}{\sqrt{Z_\pi}g_\pi}.
\end{equation}
Using (\ref{la:17}) and (\ref{la:11}), we arrive at
\begin{equation}
\label{ce:3}
f_\pi^2=4Z_\pi^{-1}\delta m^2I_2(m^2_\pi ).
\end{equation}
Hence, at $\widehat{m}\rightarrow 0$ we get
\begin{equation}
\label{ce:4}
f^2=4\!\stackrel{\circ}{\delta}\mch^2\stackrel{\circ}{I}_2,
\end{equation}
which is non--vanishing in the chiral limit.
Here and below we use the following notation
\begin{equation}
\label{ce:5}
I_2=I_2(0),\quad\stackrel{\circ}{I}_2=\lim_{\widehat{m}\rightarrow 0}I_2.
\end{equation}
\begin{equation}
\label{ce:6}
\quad\delta =1-2m\beta_\pi ,\quad\stackrel{\circ}{\delta}=\lim_{\widehat{m}
\rightarrow 0}\delta .
\end{equation}
All these quantities are convenient to simplify the formulae. Note that
the equality
\begin{equation}
\label{ce:7}
1-4G_Vf^2=\stackrel{\circ}{\delta}
\end{equation}
is valid. We shall employ the gap equation (\ref{la:9}) to expand the
constituent quark mass in a series in powers of the current
quark mass,\footnote{We remark that only analytic terms in the current
quark masses can appear within the Hartree approximation employed here.}
\begin{equation}
\label{ce:8}
m=\sum_{i=0}^{\infty}c^{(m)}_i\widehat{m}^i .
\end{equation}
Here
\begin{equation}
\label{ce:9}
c_0^{(m)}=\mch =\lim_{\widehat{m}\rightarrow 0} m,\qquad
c_i^{(m)}=\lim_{\widehat{m}\rightarrow 0} \frac{1}{i!}
          \frac{\partial^i m}{\partial\widehat{m}^i}.
\end{equation}
The mass $\mch$ is a solution of the gap equation at $\widehat{m}=0$. To
find other coefficients of the series we differentiate equation
(\ref{la:9}) with respect to the current quark mass $\widehat{m}$. As a
result, we get the following expression
\begin{equation}
\label{ce:10}
4G_{S}M^2(m)=\frac{1}{m'}-\frac{\widehat{m}}{m}.
\end{equation}
We introduced the abbreviation $M^2(m)=4m^2I_2$. In the case of exact chiral
symmetry with $\widehat{m}=0$ one can derive the value of $M^2(\mch )$ from
equation (68). Bearing in mind these remarks, we get the following
relation from the above equation:
\begin{equation}
\label{ce:11}
c_1^{(m)}=\frac{\Do}{4G_{S}f^2}.
\end{equation}
Other coefficients of the series are calculated through successive
differentiation of the gap equation. For example, differentiating equation
(74) we get
\begin{equation}
\label{ce:12}
8G_SMM'=\frac{\widehat{m}m'}{m^2}-\frac{1}{m}-\frac{m''}{(m')^2}.
\end{equation}
On the other hand, relying on the definition of $M$ one has
\begin{equation}
\label{ce:13}
\frac{MM'}{mm'}=\frac{M^2}{m^2}-\frac{3h_1}{4\pi^2},
\qquad h_1=\left(\frac{\Lambda^2}{\Lambda^2+m^2}\right)^2.
\end{equation}
In the chiral limit it follows from these two equations that
\begin{equation}
\label{ce:14}
c_2^{(m)}=\frac{3(c_1^{(m)})^2}{2\!\mch}\left(
          \frac{\mch^2\nh\stackrel{\circ}{\delta}}
          {2\pi^2f^2}-1\right).
\end{equation}
For simplicity we use the symbol $\nh$ to denote the chiral limit
\begin{equation}
\label{ce:15}
\nh =\lim_{\widehat{m}\rightarrow 0}h_1=\Lch .
\end{equation}
Thus we obtain
\begin{equation}
\label{ce:16}
  m=\mch\left[1+\frac{\mpio^2\stackrel{\circ}{\delta}}{4\mch^2}
    -\frac{3\mpio^4\stackrel{\circ}{\delta}^2}{32\mch^4}
    \left(1-\frac{\mch^2\nh\stackrel{\circ}{\delta}}{2\pi^2f^2}
    \right)+{\cal O}(\mpio^6 )\right].
\end{equation}

Another important example is the chiral expansion for the pion mass. While in
the case of the quark mass we employed the gap equation, here we need a
pion mass equation (\ref{la:19}), which can be conveniently represented as
\begin{equation}
\label{ce:17}
\left(m^2_\pi -4m\widehat{m}\frac{G_V}{G_S}\right)I_2(m^2_\pi )=
\frac{\widehat{m}}{4mG_S}.
\end{equation}
It follows from the equation that this kind of expansion begins with a
term proportional to $\widehat{m}$. We have already called it
$\stackrel{\circ}{m}^2_\pi$ (see (65)). Let us find the first few
coefficients of the series
\begin{equation}
\label{ce:18}
m^2_\pi =\sum_{i=1}^{\infty}c^{(m_\pi )}_i\widehat{m}^i.
\end{equation}
Obviously, $c^{(m_\pi )}_1=\mch\! (G_Sf^2)^{-1}$. To find other coefficients
of the series we represent equation (81) as
\begin{equation}
\label{ce:19}
4mG_S\left[\sum_{i=0}^{\infty}c^{(m_\pi )}_{i+1}\widehat{m}^i
   -4m\frac{G_V}{G_S}\right]I_2(m^2_{\pi})=1.
\end{equation}
Using the expansion (80) we isolate combinations to the same
powers of $\widehat{m}$ on the left-hand side of the equation and set them
equal to zero. Thus we can calculate coefficients of the expansion
(82). For example
\begin{equation}
\label{ce:20}
c^{(m_\pi )}_2=\frac{\Do}{(2f^2G_S)^2}\left[\frac{\mch^2\nh}{2\pi^2f^2}
               \stackrel{\circ}{\delta}(3\stackrel{\circ}{\delta}-1)+
               (1-2\stackrel{\circ}{\delta})\right].
\end{equation}
\begin{eqnarray}
\label{ce:21}
&&c^{(m_\pi )}_3=\frac{\Do^2}{4\!\mch\! (2f^2G_S)^3}\left\{
    8\!\stackrel{\circ}{\delta}\! -3+\frac{\mch^2\nh}{2\pi^2f^2}
    \left[\frac{\stackrel{\circ}{\delta}\! (2-10\!\stackrel{\circ}{\delta}
   +21\!\stackrel{\circ}{\delta}^2)\!\mch^2\nh}{2\pi^2f^2}\right.\right.
   \nonumber \\
&&\qquad\ \ \ \ \ \ \left.\left.
   +7\!\stackrel{\circ}{\delta}\!-20\!\stackrel{\circ}{\delta}^2\!\!
   -\frac{4}{5}-\frac{4(15\!\stackrel{\circ}{\delta}^2\!\!
   -10\!\stackrel{\circ}{\delta}\! +2)\!\mch^2}
   {5(\Lambda^2+\mch^2)}\right]\right\}.
\end{eqnarray}

Below we give the results of similar calculations for the pion decay
constant $f_\pi$ on the basis of the expression (\ref{ce:2}).
\begin{equation}
\label{ce:22}
f_\pi =\sum_{i=0}^{\infty}c^{(f_\pi )}_i\widehat{m}^i.
\end{equation}
In this case the corresponding coefficients of the chiral series are
\begin{equation}
\label{ce:23}
c^{(f_\pi )}_0=f,
\end{equation}
\begin{equation}
\label{ce:24}
c^{(f_\pi )}_1=\frac{\Do^2}{4\!\mch\! fG_S}\left[1-
   \frac{3\mch^2\nh\stackrel{\circ}{\delta}}{(2\pi f)^2}\right],
\end{equation}
\begin{eqnarray}
\label{ce:25}
& &c^{(f_\pi )}_2=-\frac{3\!\stackrel{\circ}{\delta}^3\!\!
   (2-\!\stackrel{\circ}{\delta})}{2f(4\!\mch\! fG_S)^2}+
   \frac{\stackrel{\circ}{\delta}^2\nh}{2f(4\pi f^2G_S)^2}
   \left\{3\!\stackrel{\circ}{\delta}^2\!\left[\frac{\Lambda^2+2\!\mch^2}
   {\Lambda^2+\!\mch^2}\right.\right.\nonumber \\
& &\left.\left.+\frac{3}{2}(1-\!\Do )
   -\frac{9\!\mch^2\nh\Do}{(4\pi f)^2}(2-\!\Do )\right]
   +\frac{\mch^2\nh\stackrel{\circ}{\delta}}{(2\pi f)^2}-
   \frac{\Lambda^2+3\!\mch^2}{5(\Lambda^2+\!\mch^2)}\right\}.
\end{eqnarray}

Concluding the section we point out that chiral expansions derived here
for the main meson characteristics prove to be helpful in establishing
correspondence between the results obtained here and the known low-energy
theorems of current algebra. We shall use them to analyze $\pi\pi$
scattering and to consider the scalar radius of the pion.

\section{$\pi\pi$-scattering}
The formal scheme described in the previous section gives the
possibility of evaluating any mesonic N-point function through the
parameters of the model: $\Lambda , \widehat{m}, G_S, G_V$. Let us
use it for the evaluation of the $\pi\pi$-scattering amplitude. We
have done that in the same approach earlier \cite{2}, but without
considering the spin-one mesons. Now we would like to explore the role
of vector and axial-vector particles in this process.

The model with linearly realized chiral symmetry was already employed for
this purpose \cite{15}.  However, the results of \cite{15} should be
considered only as a first approximation to this problem. The $\pi -a_1$
mixing was neglected there. This breaks chiral symmetry, which is
recovered only if the constituent quark mass $m$ goes to infinity. In our
case we exactly reproduce Weinberg's result at the level $E^2$ and we also
evaluate all higher order corrections in $E^2$.

The extended NJL model with nonlinearly realized chiral symmetry has not
been used to study low-energy $\pi\pi$ scattering so far, though some
general conclusions can be drawn from the results of \cite{14}. Chiral
expansion parameters $L_i$ were calculated in the model, the standard
heat kernel method being used to get the first terms of the effective
meson Lagrangian (up-to-and-including the ${\cal O}(E^4)$ terms). We shall
advance farther and find a general (to any accuracy in $E^2$) form of the
$\pi\pi$ scattering amplitude in the principal approximation in $1/N_c$.

In our calculations we use the conventional Mandelstam variables:
\begin{equation}
\label{a:1}
s=(q_1+q_2)^2,\quad t=(q_1-q_3)^2,\quad u=(q_1-q_4)^2
\end{equation}
for the scattering process $\pi^a(q_1)+\pi^b(q_2)\rightarrow
\pi^c(q_3)+\pi^d(q_4)$. The $\pi\pi$ scattering amplitude $T_{ab;cd}$
has the following isotopic structure
\begin{equation}
\label{a:2}
T_{ab;cd}(s,t,u)=\delta_{ab}\delta_{cd}A(s,t,u)+
                 \delta_{ac}\delta_{bd}A(t,s,u)+
                 \delta_{ad}\delta_{cb}A(u,t,s).
\end{equation}
It follows that the amplitudes with definite isospin are
\begin{eqnarray}
\label{a:3}
&&T^0(s,t,u)=A(t,s,u)+A(u,t,s)+3A(s,t,u), \nonumber \\
&&T^1(s,t,u)=A(t,s,u)-A(u,t,s), \nonumber \\
&&T^2(s,t,u)=A(t,s,u)+A(u,t,s).
\end{eqnarray}
Meson tree diagrams corresponding to the amplitude $A(s,t,u)$ are of the
same form both in the linear and nonlinear approach (see Fig.1). Only the
internal structure of quark-loop-based meson vertices will be different.

\subsection{The amplitude (linear approach)}
Let us calculate the amplitude $A(s,t,u)$ in the linear approach. The
vector meson and scalar meson exchange diagrams of Fig.1 include the
triangular vertices $\rho\rightarrow\pi\pi$ and $\sigma\rightarrow\pi\pi$.
We calculate these vertices in accordance with the diagrams of Fig.2,
taking into account the $\pi a_1$ mixing. The $\rho^a_\mu
(p)\rightarrow\pi^b(p_1) \pi^c(p_2)$ amplitude is equal to
\begin{equation}
\label{a:4}
M_{\rho\pi\pi}=\frac{1}{4}\mbox{Tr}(\tau_a[\tau_b,
   \tau_c]_-)(p_1-p_2)^\mu\varepsilon_\mu (p)
   f_{\rho\pi\pi}(p_1,p_2),
\end{equation}
where
\begin{equation}
\label{a:5}
f_{\rho\pi\pi}(p_1,p_2)=\frac{g_\rho (p^2)}{\sqrt{Z_{\rho}}}F(p_1,p_2).
\end{equation}
The function $F(p_1,p_2)$ has the form (for on-shell pions)
\begin{eqnarray}
\label{a:6}
&&F(p^2)=\frac{1}{Z_\pi}\biggl\{1-\frac{m\beta_\pi p^2}{
       m^2_\rho (p^2)}+\frac{\delta}{p^2-4m^2_\pi}
       \biggl[(p^2-2m^2_\pi )\left(\frac{I_2(p^2)}{I_2(m^2_\pi )}-1
       \right)     \nonumber \\
&&\ \ \ \ \ \ \ \ \ \ \ \ \ \ \ \ \ \ \ \ \ \ \ \ \ \ \ \
       +2m^4_\pi\frac{I_3(-p_1, p_2)}{I_2(m^2_\pi )}
       \biggr]\biggr\} .
\end{eqnarray}
Here $p=p_1+p_2$, with $p_1, p_2$ being the pion momenta. This function
obeys the requirement of universality of electromagnetic interactions,
$F(0)=1$. To see this one has to use the equality
\begin{equation}
\label{a:7}
I_2(0)-I_2(m^2_\pi )-m^2_\pi I_3(-p_1, p_2)\big\vert_{p^2=0}=
      2m^2_\pi\frac{\partial I_2(p^2)}{\partial p^2}
      \bigg\vert_{p^2=m^2_{\pi}}.
\end{equation}

The $\sigma^{a}(p)\rightarrow\pi^{b}(p_1)\pi^{c}(p_2)$
amplitude has the form
\begin{equation}
\label{a:8}
M_{\sigma\pi\pi}(p_1,p_2)=\frac{1}{4}\mbox{Tr}\left(\tau_{a}
   [\tau_{b},\tau_{c}]_+\right)f_{\sigma\pi\pi}(p_1,p_2).
\end{equation}
The vertex function $f_{\sigma\pi\pi}(p_1,p_2)$ on the pion mass shell is
\begin{eqnarray}
\label{a:9}
&&f_{\sigma\pi\pi}(p^2)=\frac{16mg^2_{\pi}g(p^2)}
  {Z_{\pi}\sqrt{Z_{\sigma}}}\left\{\left[1\! -\!\frac{p^2}{4m^2}(1-\delta^2)
                                   \right]\!
  I_2(p^2)+\frac{m^2_\pi\delta}{2m^2}(1-\delta )I_2(m^2_\pi )
  \right.\nonumber \\
&&\left.\ \ \ \ \ \ \ \ \ \ \ \ \ \ \ \ \ \ \ \ \ \ \ \ \ \ \ \ \ \ \ \ \ \
  +\frac{\delta^2}{2}(p^2-2m^2_\pi )I_3(-p_1,p_2)\right\}.
\end{eqnarray}

Using these expressions we obtain the contributions of the meson-exchange
diagrams in Fig.1 to the amplitude $A(s,t,u)$,
\begin{equation}
\label{a:10}
A_{\rho}(s,t,u)=4G_V\left[\frac{(s-u)F^2(t)}{1-\frac{8}{3}G_VtJ_2(t)}+
                      \frac{(s-t)F^2(u)}{1-\frac{8}{3}G_VuJ_2(u)}\right].
\end{equation}
\begin{equation}
\label{a:11}
A_{\sigma}(s,t,u)=\frac{Z_\sigma f^2_{\sigma\pi\pi}(s)}
                            {m^2_\sigma (s)-s}.
\end{equation}

Now let us consider the remaining diagrams, the boxes. Their total
number is 16. In Fig. 3 we display only structurally different diagrams.
Their calculation is quite cumbersome. Here we collect only the final result,
\begin{eqnarray}
\label{a:12}
&&A_{rest}(s,t,u)=4g^4_\pi Z^{-2}_\pi
     \left\{4[s\beta^2_\pi I_2(m^2_\pi )-I_2(s)]
     +(1-\delta^2)^2\frac{s}{m^2}
     \right.\nonumber \\
&&\times [I_2(s)-I_2(m^2_\pi )]
     +\frac{2}{3}\beta^4_\pi [(s-u)tJ_2(t)+(s-t)uJ_2(u)]
                    \nonumber \\
&&-\left. \delta^2[4(s-2m^2_\pi )I_3(q_1,-q_2)
     -4\beta^2_\pi C(s,t,u)+\delta^2B(s,t,u)]\right\},
\end{eqnarray}
where
\begin{eqnarray}
\label{a:13}
&&\!\!\!\!\!
   B(s,t,u)=(2m^4_\pi -us)I_4(q_1,q_1+q_2,q_4)
           +(2m^4_\pi -ts)I_4(q_1,q_1+q_2,q_3)\nonumber \\
&&\ \ \ \ \ \
  -(2m^4_\pi -ut)I_4(q_1,q_1-q_3,q_4)].
\end{eqnarray}
\begin{equation}
\label{a:14}
C(s,t,u)=\left(\frac{s-u}{s+u}\right)\!\left\{(t-2m^2_\pi )
         [I_2(t)-I_2(m^2_\pi )]+2m^4_\pi I_3(q_1,q_3)\right\}
        +(t\leftrightarrow u).
\end{equation}

Let us consider the low-energy expansion of the total $\pi\pi$-scattering
amplitude
\begin{equation}
\label{a:15}
A(s,t,u)=A_{\rho}(s,t,u)+A_{\sigma}(s,t,u)+A_{rest}(s,t,u).
\end{equation}
The self-consistent procedure to do that is  chiral
perturbation theory \cite{21}. The chiral expansion in powers of
external momenta and current quark masses in the case of the
$\pi\pi$ scattering amplitude has to lead to Weinberg's celebrated
theorem \cite{16} at the $E^2$ level,
\begin{equation}
\label{lee:1}
A(s,t,u)=\frac{s-\stackrel{\circ}{m}^2_\pi}{f^2}+{\cal O}(E^4),
\end{equation}
where ${\cal O}(E^4)$ is the short form for the terms of order $q^4,\
\stackrel{\circ}{m}^2_{\pi}\! q^2,\ \stackrel{\circ}{m}^4_\pi$
and higher. In the case at hand the simplest way to get this result is
to consider first of all the momentum expansion and only partly use the
quark mass expansion. We shall do a chiral expansion for the pion mass,
$m_\pi$, and the constituent quark mass, $m$, only at the last stage. In
accordance with these remarks one obtains
\begin{equation}
\label{lee:2}
A_{\rho}=\frac{4g^4_\pi}{Z^2_\pi}
    \left(\frac{3s-4m^2_\pi}{m^2}
        \right)\!\delta (1-\delta )I_2+{\cal O}(E^4).
\end{equation}
\begin{eqnarray}
\label{lee:3}
&&A_{\sigma}=\frac{4g^4_\pi}{Z^2_\pi}
      \left\{4I_2 +\frac{I_2}{m^2}
        \left[s\left(2\delta^2\! -1\right)-m^2_\pi\delta (4\delta\! -3)
        \right]\right.\nonumber \\
&&\ \ \ \ \ \ \ \ \ \ \ \
  \left. +\frac{h_1}{8\pi^2 m^2}\left[s-3\delta^2(s-2m^2_\pi )\right]
      \right\}+{\cal O}(E^4).
\end{eqnarray}
\begin{equation}
\label{lee:4}
A_{rest}=-\frac{4g^4_\pi}{Z^2_\pi}
      \left\{4I_2(1-\beta^2_\pi s)
          +\frac{h_1}{8\pi^2 m^2}\left[s-3\delta^2(s-2m^2_\pi )\right]
      \right\}+{\cal O}(E^4).
\end{equation}
Some remarks are useful here. We use momentum expansions of
integrals $I_2(s)=I_2+(sh_1)/(32\pi^2 m^2)+\ldots ,\
I_3(q_1,q_2)=-3h_1/(32\pi^2 m^2)+\ldots$. One can see that not only
divergent terms $(\sim I_2)$ contribute at this level. There are finite
contributions which are proportional to $h_1$. Full cancellation of
these terms in the full amplitude
\begin{equation}
\label{lee:5}
A(s,t,u)=\frac{4g^4_\pi\delta I_2}{Z^2_\pi m^2}(s-m^2_\pi )+{\cal O}(E^4)
\end{equation}
is a good test of self-consistency of our approach. In order to preserve
chiral symmetry they must start only at the next-to-leading order.
Noting that
\begin{equation}
\label{lee:6}
\frac{4g^4_\pi\delta I_2}{Z^2_\pi m^2}=\frac{I_2}{Z^3_\pi f^2_\pi
      I_2(m^2_\pi )}=\frac{1}{f^2}+{\cal O}(E^2),
\end{equation}
we have full agreement of our amplitude with the low-energy theorem
(105).

Now let us consider the $E^4$ correction to  Weinberg's result. After some
lengthy calculations one obtains
\begin{equation}
\label{lee:7}
A^{(4)}(s,t,u)=\frac{1}{96\pi^2f^4}\left\{2\bar{\it l}_1(s-2\!\mpio^2)^2
   +\bar{\it l}_2\left[s^2+(t-u)^2\right]\right\}.
\end{equation}
Here
\begin{equation}
\label{lee:8}
\bar{{\it l}}_1=\frac{12\pi^2f^2}{\mch^2}\stackrel{\circ}{\delta}^2\!
  \left[\stackrel{\circ}{\delta}-\frac{\mch^2\nh}{\pi^2f^2}+
  \frac{9\!\stackrel{\circ}{\delta}^3\mch^4\nh^2}{(2\pi f)^4}\right]
  -9\!\stackrel{\circ}{\delta}^4\stackrel{\circ}{h}_2-
  \frac{4\pi^2f^2}{\mch^2\stackrel{\circ}{\delta}}(1-\stackrel{\circ}
  {\delta}^2)^2,
\end{equation}
\begin{equation}
\label{lee:9}
\!\!\!\!\!\!\!\!\!
\bar{\it l}_2=3\!\stackrel{\circ}{\delta}^4\stackrel{\circ}{h}_2
  +\frac{2\pi^2f^2}{\mch^2\stackrel{\circ}{\delta}}
  (1-\!\stackrel{\circ}{\delta}^2)^2+6\!\stackrel{\circ}{\delta}^2\!
  (1-\!\stackrel{\circ}{\delta}^2)\nh ,
\end{equation}
and
\begin{equation}
\label{lee:10}
\stackrel{\circ}{h}_2=\left(\frac{\Lambda ^2
         +3\mch^2}{\Lambda ^2 +\mch^2}\right)\!\stackrel{\circ}{h}_1.
\end{equation}
Let us compare our result (111) with the known estimations of the
${\cal O}(E^4)$ terms. One can see that the structure of the term
$A^{(4)}(s,t,u)$ fully agree with the early result of \cite{22}.
There parameters $\bar{\it l}_1$ and $\bar{\it l}_2$ in the limiting case,
when $G_V\rightarrow 0$, i.e. $\stackrel{\circ}{\delta}\rightarrow 1$,
go over into the known result \cite{20}, \cite{2}.
In the case of $SU(3)$ symmetry this set of parameters are known as $L_i$.
They have been obtained in the paper \cite{14} on the basis of $SU(3)\otimes
SU(3)$ NJL Lagrangian with spin-one mesons. The constants $L_i$ specify
the general effective meson Lagrangian at order of $E^4$. One should,
however, not compare them directly to those analyzed by Gasser and
Leutwyler since their analysis includes the effect of meson loops\footnote{
And thus leads to a scale-dependence not present in the  NJL model.}
If one considers the
$SU(2)$ limit of the effective meson Lagrangian from \cite{14} and compare it
with the known $SU(2)$ effective meson Lagrangian from \cite{20} one can
relate the $SU(3)$ low-energy parameters $L_i$ with the $SU(2)$ constants
$\bar{\it l}_i$. In particular we have
\begin{equation}
\label{n:4}
L_2=2L_1=\frac{\bar{\it l}_2}{192\pi^2},\quad
L_3=\frac{\bar{\it l}_1-\bar{\it l}_2}{192\pi^2}.
\end{equation}
Or
\begin{equation}
\label{n:5}
L_2=2L_1=\frac{1}{64\pi^2}\left[\stackrel{\circ}{\delta}^4
  \stackrel{\circ}{h}_2\! +\frac{2\pi^2f^2}{3\!\stackrel{\circ}{\delta}
  \mch^2}(1-\stackrel{\circ}{\delta}^2)^2
  +2\!\stackrel{\circ}{\delta}^2\!
  (1-\stackrel{\circ}{\delta}^2)\nh\right].
\end{equation}
\begin{eqnarray}
\label{n:6}
&&L_3=\frac{1}{64\pi^2}\left[-4\!\stackrel{\circ}{\delta}^4
  \stackrel{\circ}{h}_2\! -\frac{2\pi^2f^2}{\stackrel{\circ}{\delta}
  \mch^2}(1-\stackrel{\circ}{\delta}^2)^2+2\!\stackrel{\circ}{\delta}^2
  \!(4\!\stackrel{\circ}{\delta}^2-3)\nh
  \right. \nonumber \\
&&\ \ \ \ \ \ \ \ \ \ \ \ \ \ \ \ \left.
  +\frac{4\pi^2f^2\!\stackrel{\circ}{\delta}^3}{\mch^2}\left(1-
  \frac{3\!\stackrel{\circ}{\delta}\mch^2\nh}{4\pi^2f^2}\right)^2\right].
\end{eqnarray}
To compare our formulae with similar expressions derived in \cite{14}, we
point out the following rules of correspondence:
\begin{equation}
\label{n:7}
  \Gamma (0,x)\sim\frac{4\pi^2f^2}{3\!\stackrel{\circ}{\delta}\mch^2},
  \quad\Gamma (1,x)\sim\nh ,\quad\Gamma (2,x)\sim\stackrel{\circ}{h}_2,
  \quad g_A\sim\stackrel{\circ}{\delta}.
\end{equation}
Here, we give the notation of \cite{14} on the left side and our equivalents
on the right side. Now it is easy to make sure that this part of our
calculations yields results which agree fully with earlier estimates.

Let us calculate scattering lengths $a^I_l$ and effective range
parameters $b^I_l$ up to and including ${\cal O}(E^4)$ terms. The result
is given in the Appendix since it is not new. The subsequent numerical
estimates will reveal that the $E^4$ approximation of the extended NJL
model is a very good approximation to the Hartree solution.

\subsection{The amplitude (non-linear approach)}
Let us calculate the $\pi\pi$ scattering amplitude in the nonlinear
approach. Though we used the same initial Lagrangian, the scheme developed
differs from the approach with linearly realized chiral symmetry. The
difference arises from variation in the structure of vertices describing the
interaction of Goldstone particles with quarks. Note that the structure of
similar vertices for scalar, vector, and axial-vector fields remains
unchanged. As before, we begin with the amplitude $\rho^a_\mu
(p)\rightarrow\pi^b (p_1)\pi^c(p_2)$ (see (93)). To distinguish
linear results from their nonlinear counterparts, we shall use symbols
with a tilde above them. For example, after calculation of the diagrams in
Fig. 4, instead of the function $F(p^2)$ (see (95)) we get a new
expression $\tilde{F}(p^2)$:
\begin{eqnarray}
\label{npp:1}
&&\tilde{F}(p^2)=\frac{4g^2_\pi}{Z_\pi (m-\widehat{m})^2}
  \left\{\frac{p^2}{12}(1-g_A^2)J_2(p^2)-\widehat{m}(mg_A-\widehat{m})
  I_2(m_\pi^2)
   \right. \nonumber \\
&&\!\!\!\!\!\!\!\!\!\! \left. +\frac{(mg_A-\widehat{m})^2}{p^2-4m^2_\pi}
  \left[(p^2-2m^2_\pi )\left(I_2(p^2)-I_2(m^2_\pi )\right)
  +2m^4_\pi I_3(-p_1,p_2)\right]\right\}.
\end{eqnarray}
The constant $g_A$ is equal to $1-2\tilde{\beta}_\pi F$. It can be related
to the known constant $\delta$: $(mg_A-\widehat{m})=(m-\widehat{m})
\delta$. If $\widehat{m}, p^2\rightarrow 0$, then
$\tilde{F}(p^2)\rightarrow 0$. This off-shell behaviour of the form factor
essentially differs from the behaviour of $F(p^2)$, where $F(0)=1$. The
contribution to the $\pi\pi$ scattering amplitude from the $\rho$ meson
exchange diagram will be described by formula (99), where one
should make a change $F(p^2)\rightarrow\tilde{F}(p^2)$. We call this
contribution  $\tilde{A}_\rho (s,t,u)$. For the given amplitude,
owing to the above property of the function $\tilde{F}(p^2)$, the chiral
expansion begins only with the terms ${\cal O}(E^6)$.

The vertex function for the two-pion decay of a scalar
particle\footnote{Similarly to the linear case, we shall use the symbol
$\sigma$ for this particle, though in the Lagrangian the symbol $s$ was
used.} will also differ from the linear one. It
is described by the second group of diagrams in Fig. 4.
\begin{eqnarray}
\label{npp:2}
&&\tilde{f}_{\sigma\pi\pi}(p^2)=\frac{4g(p^2)g^2_\pi}{Z_\pi
   \sqrt{Z_\sigma}(m-\widehat{m})^2 }
   \left\{(1-2g_A)(mg_A-\widehat{m})m^2_\pi I_2(m^2_\pi )
  \right. \nonumber \\
&&\ \ \ \ \ \ \ \ \ \
  +\left[mg^2_A+\widehat{m}(1-2g_A)\right]p^2I_2(p^2)
         -4m\widehat{m}(m-\widehat{m})I_2(p^2) \nonumber \\
&&\ \ \ \ \ \ \ \ \ \
  \left. +2m(mg_A-\widehat{m})^2(p^2-2m^2_\pi )I_3(-p_1, p_2)
  \right\}.
\end{eqnarray}
The form factor $\tilde{f}_{\sigma\pi\pi}(p^2)$ also tends to zero at
$\widehat{m}, p^2\rightarrow 0$. Changing $f_{\sigma\pi\pi}$ for
$\tilde{f}_{\sigma\pi\pi}$ in (100) we get the contribution to the
$\pi\pi$ scattering amplitude from the exchange of the given scalar. We
denote it by $\tilde{A}_\sigma (s,t,u)$. The chiral expansion of
$\tilde{A}_\sigma (s,t,u)$ begins with terms of order
${\cal O}(E^4)$.

Now let us consider the remaining group of diagrams displayed in Fig. 5.
Here we do not show diagrams resulting from elimination of $\pi a_1$
mixing. These diagrams are easily derived from the given ones by the
corresponding inserts in vertices $\xi$ with one pion at the end. Without
going into detail, we shall give the total contribution to the $\pi\pi$
scattering amplitude from all the diagrams taken together,
\begin{eqnarray}
\label{npp:3}
&&\tilde{A}_{rest}(s,t,u)=\frac{4g^4_\pi}{Z_\pi^2(m-\widehat{m})^4}
  \left\{(mg_A-\widehat{m})\left[(m-3\widehat{m})(s-m^2_\pi )
  \right.\right. \nonumber \\
&&\left. +2\widehat{m}(1-2g_A)m^2_\pi\right]\!I_2(m^2_\pi )
  +\frac{(1-g_A^2)^2}{24}\left[(s-u)tJ_2(t)+(s-t)uJ_2(u)\right]
   \nonumber \\
&&+\widehat{m}\left[\widehat{m}(1-g_A)^2s+2g_A(mg_A-\widehat{m})s
  -4\widehat{m}(m-\widehat{m})^2\right]I_2(s)
  \nonumber \\
&&+g^2_A(mg_A-\widehat{m})(mg_A-3\widehat{m})s
  \left[I_2(s)-I_2(m^2_\pi )\right]
  \nonumber \\
&&+(mg_A-\widehat{m})^2
   \left[4\widehat{m}(m-\widehat{m})(s-2m^2_\pi )I_3(q_1, -q_2)
   \right. \nonumber \\
&& \left.\left. -(1-g^2_A)C(s,t,u)-(mg_A-\widehat{m})^2B(s,t,u)
   \right]\right\}.
\end{eqnarray}
Here we use the notation (102) and (103) adopted earlier.

The expressions for the separate amplitudes $A_{\rho ,\sigma , rest}(s,t,u)$
in the linear and nonlinear approaches are different. Before we will consider
the total amplitude $A(s,t,u)=\tilde{A}_\rho +\tilde{A}_\sigma +
\tilde{A}_{rest}$ let us show that its chiral expansion coincide up to and
including the terms ${\cal O}(E^4)$ with the corresponding result in the
linear approach. Indeed, in the main ${\cal O}(E^2)$ approximation we get
Weinberg's result which follows from the term
\begin{equation}
\label{n:1}
\frac{4g^4_\pi}{Z_\pi^2(m-\widehat{m})^4}
  (mg_A-\widehat{m})(m-3\widehat{m})(s-m^2_\pi )I_2(m^2_\pi )=
  \frac{s-\mpio^2}{f^2}+\mbox{\cal O}(E^4).
\end{equation}

In the ${\cal O}(E^4)$ approximation the contribution comes from only two
amplitudes, $\tilde{A}_{\sigma}(s,t,u)$ and $\tilde{A}_{rest}(s,t,u)$.
As was established, $\rho$ meson exchange diagrams lead to an amplitude
beginning with the term of the order ${\cal O}(E^6)$. Here we get
\begin{eqnarray}
\label{n:2}
&&\tilde{A}^{(4)}_{\sigma}=\frac{8\!\mch\!\widehat{m}}{f^4}
  \left\{\left[2\!\mch\!\widehat{m}-s\!\Do^2\!\! -\!\Do (1-2\!\Do )\!\mpio^2
  \right]\!\Io +\frac{3\!\Do^2\nh}{16\pi^2}(s-2\!\mpio^2)\right\}
  \nonumber \\
&&\ \ \ \ \ \ \ \
  +\frac{\Do}{4\!\mch^2\!\! f^2}\left[\Do\! (s-2\!\mpio^2)
  \left(1-\frac{3\mch^2\Do\nh}{4\pi^2f^2}\right)+\mpio^2\right]^2,
\end{eqnarray}
\begin{eqnarray}
\label{n:3}
&&\tilde{A}^{(4)}_{rest}=-\frac{8\!\mch\!\widehat{m}}{f^4}
  \left\{\left[2\!\mch\!\widehat{m}-s\!\Do^2\!\! -\!\Do (1-2\!\Do )\!\mpio^2
  \right]\!\Io +\frac{3\!\Do^2\nh}{16\pi^2}(s-2\!\mpio^2)\right\}
  \nonumber \\
&&\ \ \ \ \ \ \ \ \ \
  +\frac{(1-\!\Do^2)^2}{6f^4}[(s-u)t+(s-t)u]\Io
  -\frac{\mpio^2\Do}{4\!\mch^2\!\! f^2}\left[2s\!\Do
  +(1-4\!\Do )\!\mpio^2\right]
  \nonumber \\
&&\ \ \ \ \ \ \ \ \ \
  +\frac{\Do^2\nh}{8\pi^2f^4}\left\{s(s-\!\mpio^2)\Do^2
  -\mpio^2\left[3s(1-\!\Do )+2\!\mpio^2(3\!\Do -2)\right]
  \right. \nonumber \\
&&\ \ \ \ \ \ \ \ \ \ \ \ \ \ \ \ \ \ \ \ \ \
  \left. +(1-\!\Do^2)[(s-u)(t+\!\mpio^2)+(s-t)(u+\!\mpio^2)]\right\}
  \nonumber \\
&&\ \ \ \ \ \ \ \ \ \
  -\frac{\Do^4\nhh}{8\pi^2f^4}\left[2\!\mpio^4\! +ut-s(t+u)\right].
\end{eqnarray}
Summing the expressions, we arrive at a result coinciding excately with
the result of similar linear calculations (111). Noteworthy is
that contributions proportional to $\widehat{m}$ are fully cancelled by
the summation. Thus, despite all the differences in the two approaches,
they are equivalent up-to-and-including
the terms of the order ${\cal O}(E^4)$ in the chiral expansion.

\subsection{Equivalence of the linear and the non-linear approach}
Haag's theorem \cite{23} in axiomatic field theory
states the independence of the $S$-matrix elements on mass-shell from
the choice of interpolating fields. In the framework of the Lagrangian
approach the same result exists \cite{24}. In our case it means in
particular that if we did all correctly the total amplitude $A(s,t,u)$
should be the same in both cases. To demonstrate it let us first consider
the $\rho\pi\pi$ form factors $F(p^2)$ and $\tilde{F}(p^2)$
(see (95) and (119)).
One then obtains
\begin{equation}
\label{h:1}
\tilde{F}(p^2)-F(p^2)=\frac{1}{Z_\pi}\left(\frac{m}{m-\widehat{m}}\right)
   \left[\frac{p^2}{m^2_{\rho}(p^2)}-1\right].
\end{equation}
It is clear now that on the $\rho$ mass shell these form factors coincide
with each other. We have a similar behaviour for $f_{\sigma\pi\pi}$
and $\tilde{f}_{\sigma\pi\pi}$
(see expressions (98) and (120)),
\begin{equation}
\label{h:2}
\tilde{f}_{\sigma\pi\pi}(p^2)-f_{\sigma\pi\pi}(p^2)=
  \frac{4g^2_\pi g(p^2)I_2(p^2)}{Z_\pi\sqrt{Z_\sigma}(m-\widehat{m})}
  \left[p^2-m^2_\sigma (p^2)\right].
\end{equation}
Again at $p^2=m^2_\sigma$ one gets zero for this difference.

We are ready now to compare the amplitudes
\begin{equation}
\label{h:3}
\Delta_\alpha (s,t,u)=\tilde{A}_\alpha (s,t,u)-A_\alpha (s,t,u),
\end{equation}
where $\alpha =\rho , \sigma , rest$. The result can be written as
\begin{eqnarray}
\label{h:4}
&&\!\!\!\!
  \Delta_\rho (s,t,u)=\frac{4g^4_\pi\delta (1-\delta )}{Z^2_\pi
            m(m-\widehat{m})}\left[\left(\frac{2\widehat{m}-m}
            {m-\widehat{m}}\right)(3s-4m^2_\pi )I_2(m^2_\pi )
  \right. \nonumber \\
&&\ \ \ \ \ \ \ \ \ \ \ \ \ \ \ \ \ \ \ \
  \left. -D(s,t,u)+2\delta C(s,t,u)\right].
\end{eqnarray}
\begin{eqnarray}
\label{h:5}
&&\!\!\!\!
  \Delta_\sigma (s,t,u)=\frac{4g^4_\pi}{Z^2_\pi (m-\widehat{m})^2}
  \left\{m^2_\pi\delta I_2(m^2_\pi )+(4m^2-s)I_2(s)
  \right. \nonumber \\
&&-4m(m-\widehat{m})\left[2\left(1-\frac{s(1-\delta^2)}{4m^2}\right)I_2(s)
  +\frac{m^2_\pi}{m^2}\delta (1-\delta )I_2(m^2_\pi )
  \right. \nonumber \\
&&\left.\left. +\delta^2(s-2m^2_\pi )I_3(q_1,-q_2)\right]\right\}.
\end{eqnarray}
\begin{eqnarray}
\label{h:6}
&&\Delta_{rest}(s,t,u)=\frac{4g^4_\pi}{Z^2_\pi}\left\{\left[
  \frac{s+4m(m-2\widehat{m})}{(m-\widehat{m})^2}-
  \frac{2s(1-\delta^2)}{m(m-\widehat{m})}\right]I_2(s)
  \right. \nonumber \\
&&+\frac{m^2_\pi\delta}{m(m-\widehat{m})^2}
   \left[4\widehat{m}(1-\!\delta\! )-m+\frac{3s}{m^2_\pi}(1-\!\delta\! )
   (m-2\widehat{m})\right]I_2(m^2_\pi )
   \nonumber \\
&&+\frac{4m\delta^2}{(m-\widehat{m})}(s-2m^2_\pi )I_3(q_1,-q_2)
   \nonumber \\
&&\left.
  +\frac{\delta(1-\delta )}{m(m-\widehat{m})}\left[D(s,t,u)-
  2\delta C(s,t,u)\right]\right\}.
\end{eqnarray}
Here the notation
\begin{equation}
\label{h:7}
D(s,t,u)=\frac{(1-\delta )}{6\delta m^2}\left(\delta +
         \frac{\widehat{m}}{m-\widehat{m}}\right)
         [(s-u)tJ_2(t)+(s-t)uJ_2(u)]
\end{equation}
has been used. Summing (128)-(130) it is easy to see that
\begin{equation}
\label{h:8}
\Delta_\rho (s,t,u)+\Delta_\sigma (s,t,u)+\Delta_{rest}(s,t,u)=0.
\end{equation}
This is what we wanted to show.

\subsection{The full result}
Now we can go on and estimate the role of higher order contributions
in the amplitude obtained. For direct comparison with the empirical
numbers we fix parameters in order to obtain the pion decay constant,
$f_\pi$, and the pion mass, $m_\pi$, close to their physical values,
$f_\pi\simeq 93$ MeV and $m_\pi\simeq 139$ MeV, respectively. We take
two sets of parameters, set $I$ with a
constituent quark mass $m \simeq m_\rho /2$, set
$II$ with a rather low quark mass. In the first case we get a bound state
for the $\rho$ - meson, in the other case the $\rho$ - meson mass lies above
the threshold for production of a pair of 'free' quarks, a problem related
to the absence of confinement of the model. Nevertheless, as will be discussed
in the next section, a better description of the scalar form factor
of the pion is only achieved for rather low constituent quark masses, so that,
with due care, it is of interest to study pion observables for this case too.
The model parameter $G_V$ is obtained by fixing the mass of the
$\rho$ meson, $m_\rho\simeq 770$ MeV, as long as $m_\rho < 2 m$. For the
low constituent mass case, the $\rho$ is embedded deeply in the $\bar q q$
continuum. To avoid the complications of defining this isovector-vector state
under such circumstances \cite{fred}, we choose to fit the scattering length
$a_1^1$ to fix $G_V$. This is an unambiguous and rather simple procedure.
 As a result, the four parameters of
the model for set $I$ have the values: $G_S=9.41\:\mbox{GeV}^{-2},\ G_V=11.29
\:\mbox{GeV}^{-2},\ m=390\:\mbox{MeV}\ (\widehat{m}=3.9\:\mbox{MeV})$ and
$\Lambda =1$ GeV. With these parameters, we find $f_\pi =92\:\mbox{MeV},
\ m_\pi =139\:\mbox{MeV},\ m_\rho =770\:\mbox{MeV},\ \delta=0.62.$
For set $II$ we obtain $G_S=1.083\:\mbox{GeV}^{-2},\ G_V=8.8\:
\mbox{GeV}^{-2},\ m=200\:\mbox{MeV}\ (\widehat{m}=1.0\:\mbox{MeV})$,
$\Lambda =2.5$ GeV, $f_\pi =92.7\:\mbox{MeV},
\ m_\pi =139\:\mbox{MeV}, \ \delta=0.69.$ For obvious reasons,
phase shifts  are always presented for $\sqrt {s} < 2m$.
We also calculate the set of parameters in the limiting case
$G_V\rightarrow 0$ or $\delta \rightarrow 1$, already  determined in \cite{2}.
Here we use $G_S=7.74\:\mbox{GeV}^{-2},\quad \Lambda =1\:\mbox{GeV}$
and $m=242\:\mbox{MeV}\quad (\widehat{m}=5.5\:\mbox{MeV})$. Therefore one can
always compare the predictions of the NJL and extended NJL models.
Apart from that, we need values of the main physical quantities in
the chiral limit $\widehat{m}\rightarrow 0$. In this
case we obtain for set $I$, $\mch =382\:\mbox{MeV},
\ f=90.9\:\mbox{MeV},\ \Do =0.628$ and for set $II$,
$\mch =183\:\mbox{MeV},
\ f=88.4\:\mbox{MeV},\ \Do =0.73$
The leading term of the chiral expansion for the pion mass is $\mpio =138.42$
MeV in $I$ and $\mpio =141$ MeV in set $II.$

First, we give the values for the constants $L_2$ and $L_3$: we have
for set $I$, $L_2=1.2\cdot 10^{-3}$, $L_3=-3.2\cdot 10^{-3}$ and for set
$II$, $L_2=2.0\cdot 10^{-3}$, $L_3=-2.2\cdot 10^{-3}$.
For comparison, we give
 the scale--dependent empirical values $L_2 (m_\rho)
 =1.2\cdot 10^{-3}$, $L_3 (m_\rho )=-3.6\cdot 10^{-3}$. However such a
 comparison has to be taken cum grano salis since the $L_i$ calculated
 within the Hartree approximation are simple c-numbers.
Whereas set $I$ lead to values closer to the experimental ones, set $II$ is
compatible with the results of \cite{14}.
Second, we consider the $\pi\pi$ threshold parameters. The results of our
calculations are presented in the Table. The first and second columns list
the numbers based on the chiral expansion ($E^4$ approximation), for sets $I$
and $II$. The formulae used are given in the Appendix. The third and fourth
columns show the results of the exact calculations on the basis of the full
amplitudes $A(s,t,u)$, for $I$ and $II$. We compare them with the results
from the standard NJL approach without vector mesons \cite{2}, within
the current algebra \cite{16}, and with the experimental data \cite{25}.

\begin{table}
\begin{tabular}{||l||l|l|l|l|l|l|l||} \hline
$a^I_l$ & ${\cal O}(E^4) [I]$ & ${\cal O}(E^4) [II]$ & total [I] & total [II]
& NJL & SMT & exp. \\ \hline
$a^0_0$ & 0.16 & 0.19 & 0.17 & 0.19 & 0.19 &0.16 &$0.26\pm 0.05$ \\
$b^0_0$ &0.18&0.25&0.19&0.25&0.27&0.18&$0.25\pm 0.03$ \\
$a^2_0$ &-0.046&-0.043&-0.047&-0.045&-0.044&-0.045&$-0.028\pm 0.012$ \\
$b^2_0$ &-0.088&-0.078&-0.090&-0.079&-0.079&-0.089&$-0.082\pm 0.008$ \\
$a^1_1$ &0.034&0.037&0.038&0.039$^*$&0.034&0.030&$0.038\pm 0.002$ \\
$a^0_2\times 10^{4}$&5.9&21.0&6.9&18.5&16.7&&$17\pm 3$ \\
$a^2_2\times 10^{4}$&-2.1&4.7&-2.5&0.0 &3.2&&$1.3\pm 3$ \\ \hline
\end{tabular}\\[0.5cm]
{\small {\bf Table:} The $\pi\pi$ scattering lengths and effective ranges
in the extended NJL model in comparison with the same calculations in the
original NJL model, without spin-one mesons \cite{2}, and the soft meson
theorems (SMT) \cite{16}. The experimental data are taken from \cite{25}.}
The '$^*$' denotes an input quantity.
\end{table}

{}From this comparison one infers the following.
 All scattering lengths and range parameters
are mainly determined by the ${\cal O}(E^4)$ approximation. The effect of
higher-order terms is noticeable only in the D-waves.
If we compare the results of numerical calculations with our
previous estimations of low-energy $\pi\pi$ scattering parameters within
the NJL model without vector mesons, we notice that the
agreement with the experimental data is of comparable quality (for set II)
but poorer for set I. This confirms the observation of Ref.\cite{27}
that within the Hartree approximation one needs to work with a small
constituent quark mass. Similar findings were obtained in Ref.\cite{14}.
Alternatively, one could think of going beyond the Hartree approximation
and, in particular, include pion loops. This is a difficult problem
which deserves further studies \cite{loops}.

In Fig.6a,b  we show the S and P-wave  phase shifts $\delta_0^0$, $\delta_0^2$
and  $\delta_1^1$ for set I in comparison with the available data ($\sqrt s
\le 700$~MeV). The calculated phases are in reasonable agreement with the data.
For set II, we are confined to $\sqrt s \le 400$~MeV and therefore only
show the S-wave phases in Fig.6c. New data in this region will come once
DA$\Phi$NE is operating and $K_{\ell 4}$ decays have been analyzed.

\section{The scalar form factor of the pion}
Another quantity closely related to $\pi\pi$ scattering is the scalar
form factor of the pion, defined by
\begin{equation}
<\pi^a (p')\mid\widehat{m} (u\bar{u}+d\bar{d})\mid \pi^b (p)>=
\delta^{ab}\Gamma_\pi (t)
\label{e:1}
\end{equation}
where $t=(p-p')^2$ is the square of the invariant four-momentum transfer.
In the framework of the original NJL model this quantity was previously
determined \cite{26} and agreed quite well with the empirical scalar form
factor of the pion for low momentum transfers at a fairly small constituent
quark mass of $241.8$ MeV. In particular, the scalar pion radius was
demonstrated to impose powerful constraints on the parameters of the NJL
model. We anticipate that with the large constituent mass as in set I, the
scalar radius will be largely underestimated.
The relevant Feynman diagrams to calculate this quantity
are depicted in Fig.7. There is a sum of
two contributions to the scalar pion form factor in the NJL model: one is
the direct coupling of the operator $\widehat{m}(u\bar{u}+d\bar{d})$
(double-dashed line) to the two pions via a quark triangle (we call it
bare coupling), and the other corresponds to rescattering of quarks into
the scalar meson $\sigma$, which in turn couples to the pions. In Fig.7a
we show these contributions, with the pion legs amputated. After solving the
Bethe--Salpeter equation for this vertex, $\Gamma$, one obtains the scalar
pion form factor by attaching the pion legs, Fig.7b. With eqs.
(20) and (98), the result for the scalar form factor of the
pion $\Gamma_\pi (t)$ can be cast in the form
\begin{equation}
\Gamma_\pi (t)=\frac{\widehat m (4g_\pi)^2}{Z_\pi}
\left[1+\frac{{\cal J}_s(t)g^2(t)}{m^2_\sigma (t)-t}\right]{\cal K}(t)
\label{e:2}
\end{equation}
where
\begin{equation}
{\cal K}(t)=\left[m-\frac{t\beta_\pi}{2}(1+\delta )\right]\! I_2(t)
           +m_\pi^2 \beta_\pi\delta I_2(m_\pi^2)
           +\frac{m}{2}\delta^2(t-2 m_\pi^2)I_3(p,p'),
\label{e:3}
\end{equation}
and ${\cal J}_s(t)=8I_1+4(t-4m^2)I_2(t)$ is the fundamental quark bubble in
the scalar channel. For the same parameter sets used in the evaluation of the
scattering parameters, we obtain $\Gamma_\pi (0)=1.007m_\pi ^2$, (set $I$),
and $\Gamma_\pi (0)=0.946m_\pi ^2$, (set $II$), consistent
with the $\chi PT$ prediction \cite{21}. By expanding in powers of
$t$ we extract the scalar mean square radius of the pion
\begin{equation}
\frac {\Gamma_\pi(t)}{\Gamma_\pi(0)} =1+\frac{1}{6}<r^2>^s_\pi t+
{\cal O}(t^2).
\label{e:4}
\end{equation}
Its numerical value $<r^2>^s_\pi =0.043\:\mbox{fm}^2$ calculated with the
parameter set $I$ is one order of magnitude smaller than the empirical
value $<r^2>^s_\pi =(0.55 \pm 0.15)\:\mbox{fm}^2$ \cite{27} extracted
from phase shift analyses \cite{28}. For set $II$ we increase the mean scalar
radius $<r^2>^s_\pi =0.53\:\mbox{fm}^2$ in good agreement with the
empirical number. To understand these numbers in more
detail, we investigate the chiral expansion of the scalar form factor
 up-to-and-including terms of  order $E^4$,
\begin{equation}
\label{e:5}
\Gamma_\pi (0)=\mpio^2\left\{1+\frac{\mpio^2\stackrel{\circ}{\delta}}
               {2\!\mch^2}\left[1-2\!\stackrel{\circ}{\delta}\!
               -\frac{\mch^2\nh\stackrel{\circ}{\delta}\!
               (1-3\!\stackrel{\circ}{\delta})}{2\pi^2f^2}
               \right]\right\}+{\cal O}(E^6).
\end{equation}
In the leading order of the chiral expansion we get exactly the current
algebra result $\Gamma_\pi (0)=\mpio^2$ \cite{29}, as expected, since all
symmetries are respected in the evaluation of the form factor. The $E^4$
contribution leads to $\Gamma_\pi (0)=1.005\mpio^2$ for set $I$ and
$\Gamma_\pi (0)=0.943\mpio^2$ in set $II.$
For the scalar radius of the pion we have
\begin{equation}
\label{e:6}
<r^2>^s_\pi =\frac{3\!\stackrel{\circ}{\delta}^2}{2\!\mch^2}
             \left[1-\frac{3\!\mch^2\stackrel{\circ}{\delta}\nh}
             {(2\pi f)^2}\right]
\end{equation}

\noindent which is $\approx 0.057\:\mbox{fm}^2$ in set $I$ and
$\approx 0.70\:\mbox{fm}^2$ in set $II$,
i.e. in this quantity one observes large
higher order effects. This is not unexpected from previous calculations
in chiral perturbation
theory beyond one loop \cite{28}. These higher order effects are essentially
accounted for by the low constituent quark mass in agreement with the
findings of Ref.\cite{27}. We stress again that alternatively one might want
to consider pion loop effects, which, however, goes beyond the scope of the
present manuscript.

\section{Summary and conclusions}
In the present paper we have developped a method which allows for the
calculation
of $N$-point functions within the extended Nambu--Jona-Lasinio model with
nonlinear realization of chiral symmetry \cite{14}. This extends
the ideas described in \cite{2}. Using the path
integral technique and the Hartree approximation, we work in
momentum space to circumvent the standard way of employing the heat kernel
expansion for constructing an effective meson Lagrangian. With the heat kernel
method, one manages to get only the first few terms of the Lagrangian
(in general the
${\cal O}(E^4)$ approximation) and faces growing difficulties when
calculating terms of yet higher orders. Examining the one-loop
approximation for the effective action in momentum space, we find
transformations of collective variables which permit to calculate these higher
order terms in a straightforward manner. The two-point functions describing
propagation of collective excitations lead to results coinciding with those
obtained from analysis of Bethe--Salpeter equations on bound
quark-antiquark states for two-particle Green functions of quark fields.

Using equations for the mass of constituent quarks (gap equation) and the
pion mass, we construct chiral expansions for these quantities and the
constant $f_\pi$. We have also calculated  the $\pi\pi$ scattering
amplitude. This
program is applied both to the linear and nonlinear description of
transformation properties of physical fields. We have demonstrated explicitely
that  the $\pi\pi$ scattering
amplitude $A(s,t,u)$ obtained in both approaches is completely equivalent.
We have carried out the chiral expansion for
this amplitude and calculated the pertinent
scattering lengths and range parameters. In addition we have obtained the
corresponding low-energy $\pi \pi$ phase shifts. All these calculations
have been performed for two sets of parameters, the first and the second
one having a large ($\sim 400$ MeV) and a small ($\sim 200$ MeV)
constituent mass, respectively.
In general, the conclusion is
that the corrections derived by considering terms of the effective meson
Lagrangian with higher derivatives are small for S--
and P--waves but are significant in the D--waves. Agreement with the
experimental data is of comparable quality than in the model ignoring vector
modes for the low constituent quark mass case. This conclusion has been
sharpened further by studying
the scalar pion radius. Only for low constituent masses one can describe this
quantity within the Hartree approximation of the (extended) NJL model
\cite{27}. Such low constituent quark masses are also obtained in the currently
popular estimates of the chiral perturbation theory low-energy constants from
extended NJL models \cite{14}. Further research in this direction should be
concerned with a consistent implementation of pion loop effects (for some first
attempts see e.g.
\cite{loops}).

\vspace{1cm}

\appendix{\bf APPENDIX}
\vspace{0.5cm}

Here we give expressions for the main low-energy characteristics of
the $\pi\pi$ scattering derived in the ${\cal O}(q^4)$ approximation
from the amplitude (105) + (111). They coincide with the
results of \cite{20}.
\begin{equation}
\label{ap:1}
a^0_0=\frac{7\!\mpio^2}{32\pi f^2}\left[1+\frac{5\mpio^2}{84\pi^2f^2}
      \left(\bar{\it l}_1+2\bar{\it l}_2-\frac{9\bar{\it l}_3}{10}
      \right)\right].
\end{equation}
\begin{equation}
\label{ap:2}
a^2_0=-\frac{\mpio^2}{16\pi f^2}\left[1-\frac{\mpio^2}{12\pi^2f^2}
      \left(\bar{\it l}_1+2\bar{\it l}_2\right)\right].
\end{equation}
\begin{equation}
\label{ap:3}
\mpio^2b^0_0=\frac{\mpio^2}{4\pi f^2}\left[1+\frac{\mpio^2}{12\pi^2f^2}
      \left(2\bar{\it l}_1+3\bar{\it l}_2\right)\right].
\end{equation}
\begin{equation}
\label{ap:4}
\mpio^2b^2_0=-\frac{\mpio^2}{8\pi f^2}\left[1-\frac{\mpio^2}{12\pi^2f^2}
      \left(\bar{\it l}_1+3\bar{\it l}_2\right)\right].
\end{equation}
\begin{equation}
\label{ap:5}
\mpio^2a^1_1=\frac{\mpio^2}{24\pi f^2}\left[1+\frac{\mpio^2}{12\pi^2f^2}
      \left(\bar{\it l}_2-\bar{\it l}_1\right)\right].
\end{equation}
\begin{equation}
\label{ap:6}
\mpio^4a^0_2=\frac{\mpio^4}{1440\pi^3f^4}\left(\bar{\it l}_1+
             4\bar{\it l}_2\right).
\end{equation}
\begin{equation}
\label{ap:7}
\mpio^4a^2_2=\frac{\mpio^4}{1440\pi^3f^4}\left(\bar{\it l}_1+
             \bar{\it l}_2\right).
\end{equation}
The quantity $\bar{\it l}_3$ appears in the chiral expansion for the pion
mass
\begin{equation}
\label{ap:8}
m^2_\pi =\mpio^2\left(1-\frac{\mpio^2\bar{\it l}_3}{32\pi^2f^2}\right).
\end{equation}
It is
\begin{equation}
\label{ap:9}
\bar{\it l}_3=4\!\Do\!\left[\Do (1-3\!\Do )\nh
             +\frac{2\pi^2f^2}{\mch^2}(2\!\Do\! -1)\right].
\end{equation}
For the pion decay constant one can obtain
\begin{equation}
\label{ap:10}
f_\pi =f\left(1+\frac{\mpio^2}{(4\pi f)^2}\bar{\it l}_4\right),
\end{equation}
where
\begin{equation}
\label{ap:11}
\bar{\it l}_4=\Do^2\!\left[\left(\frac{2\pi f}{\mch}\right)^2-3\!\Do\nh
                     \right].
\end{equation}

\baselineskip 12pt plus 2pt minus 2pt

\newpage
{\bf Figure captions}
\vspace{0.5cm}

Fig.1. One-loop quark diagrams corresponding to the amplitude $A(s,t,u)$.
(a) $\rho$ meson exchange; (b) scalar particle exchange; (c) other
contributions. The structure of meson vertices is shown in other figures.
\vspace{0.5cm}

Fig.2. Vertices $\rho\rightarrow\pi\pi$ and $\sigma\rightarrow\pi\pi$ in
the model with linear realization of chiral symmetry. \vspace{0.5cm}

Fig.3. Diagrams corresponding to vertex (c) (see Fig.1) in the model with
linear realization of chiral symmetry. \vspace{0.5cm}

Fig.4. Vertices $\rho\rightarrow\pi\pi$ and $\sigma\rightarrow\pi\pi$ in
the model with nonlinear realization of chiral symmetry. Added to
these diagrams should be effects of $\pi a_1$ mixing on lines with
$\xi_\mu$. \vspace{0.5cm}

Fig.5. Diagrams corresponding to vertex (c) (see Fig.1) in the model with
nonlinear realization of chiral symmetry. Added to these diagrams should be
effects of $\pi a_1$ mixing on lines with vertex $\xi_\mu$ (with the one
pion field only).
\vspace{0.5cm}

Fig.6a. The S-wave phase shifts for the heavy mass case (in degrees).
Upper panel: $\delta_0^0$, lower panel: $\delta_0^2$.
Data for $\delta_0^0$ are from Refs.\cite{dat00} and for $\delta_0^2$
from Refs.\cite{dat02}.
\vspace{0.5cm}

Fig.6b. The P-wave phase shift for the heavy mass case (in degrees).
Data are from Refs.\cite{dat11}. \vspace{0.5cm}

Fig.6c. The S-wave phase shifts for the low mass case (in degress).
Upper panel: $\delta_0^0$, lower panel: $\delta_0^2$.
\vspace{0.5cm}

Fig.7a. The full scalar quark form factor, $\Gamma$, is the sum of the bare
coupling of the operator $\widehat{m} (u\bar{u}+d\bar{d})$ (double line)
to the quarks and a contribution from rescattering of the quarks
through the four-fermion interaction in the scalar channel, with
strength $G_S$. The latter can be represented as a coupling to the
composite meson $\sigma$.  \vspace{0.5cm}

Fig.7b. The scalar pion form factor $\Gamma_\pi$ obtained by attaching
two pion legs to $\Gamma$.
\vspace{2cm}

\begin{figure}[bht]
\centerline{
\epsfxsize=6in
\epsfysize=8in
\epsffile{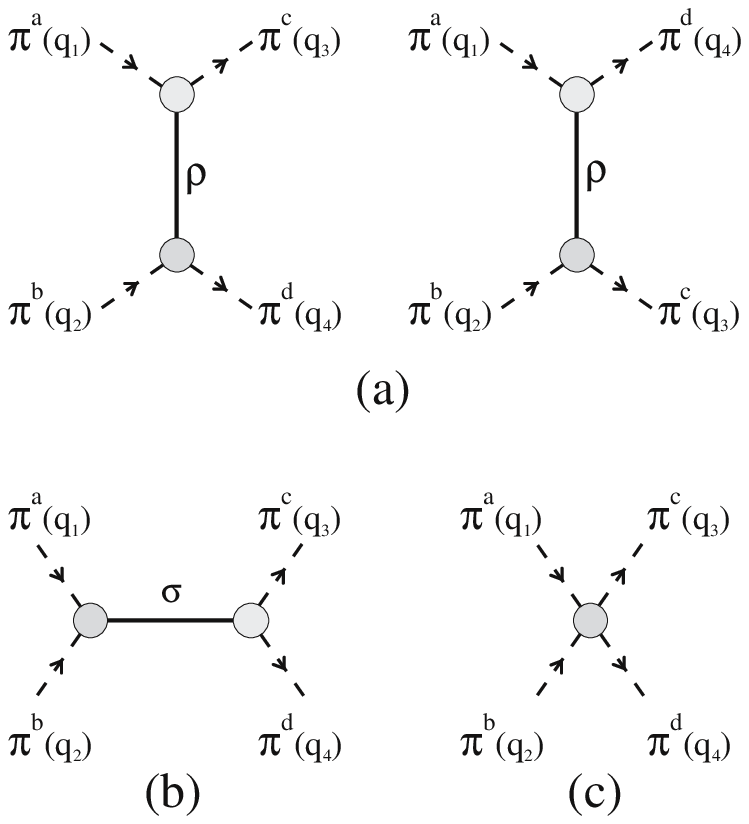}
}
\vskip -1.5cm

\centerline{\Large Figure 1}
\end{figure}


\begin{figure}[bht]
\centerline{
\epsfxsize=6in
\epsfysize=8in
\epsffile{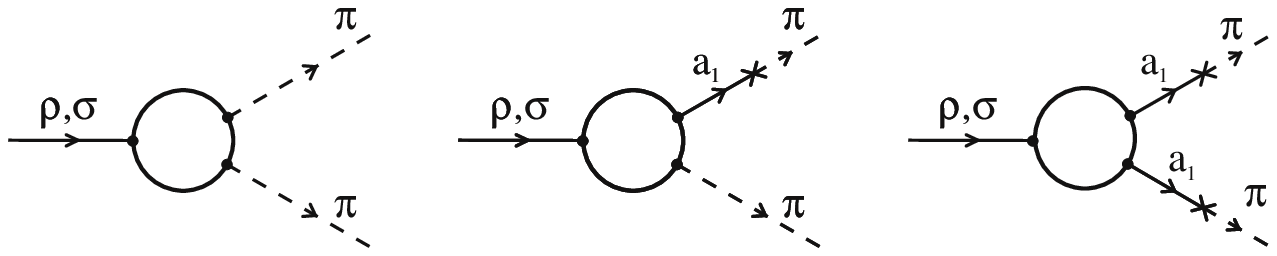}
}
\vskip -1.5cm

\centerline{\Large Figure 2}
\end{figure}


\begin{figure}[bht]
\centerline{
\epsfxsize=6in
\epsfysize=8in
\epsffile{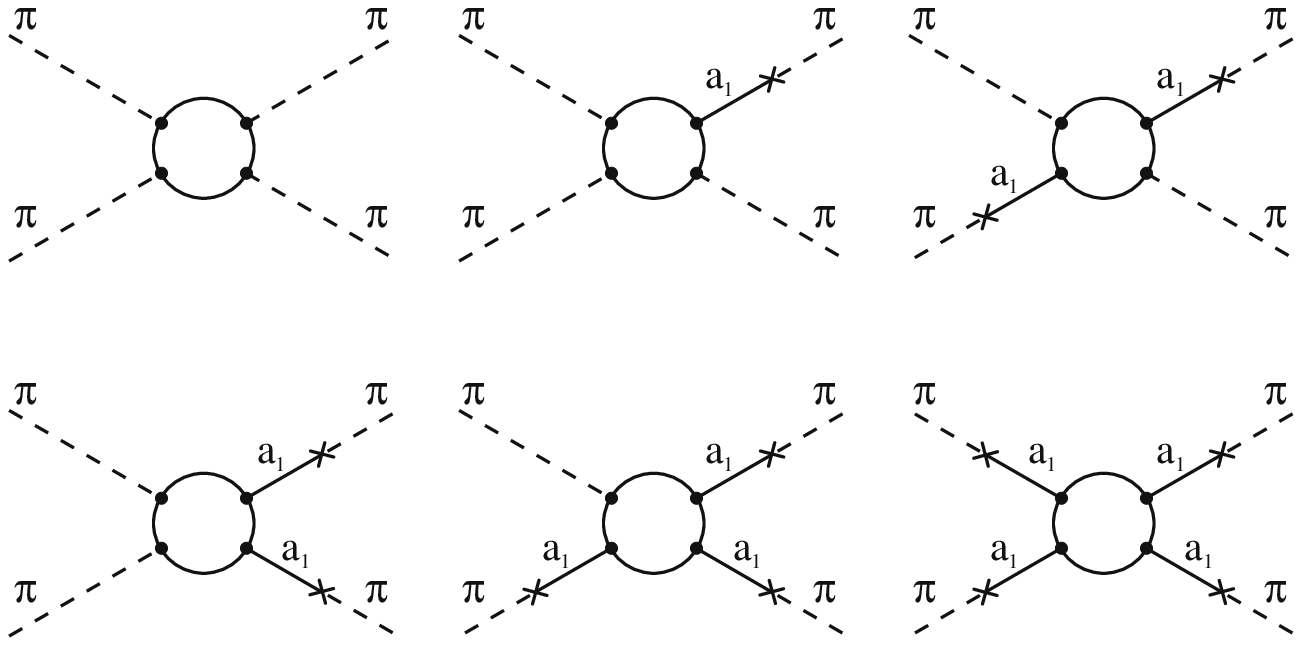}
}
\vskip -1.5cm

\centerline{\Large Figure 3}
\end{figure}


\begin{figure}[bht]
\centerline{
\epsfxsize=6in
\epsfysize=8in
\epsffile{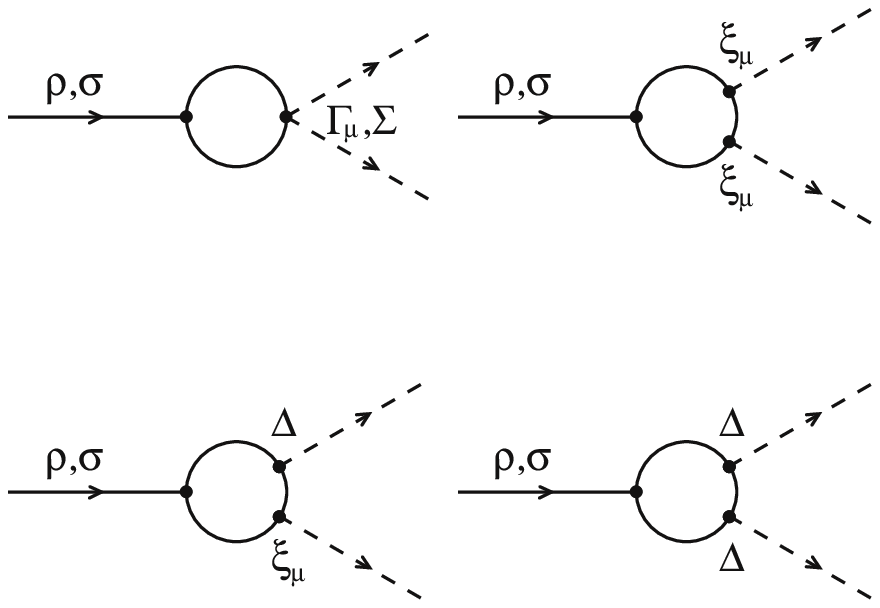}
}
\vskip -1.5cm

\centerline{\Large Figure 4}
\end{figure}


\begin{figure}[bht]
\centerline{
\epsfxsize=6in
\epsfysize=8in
\epsffile{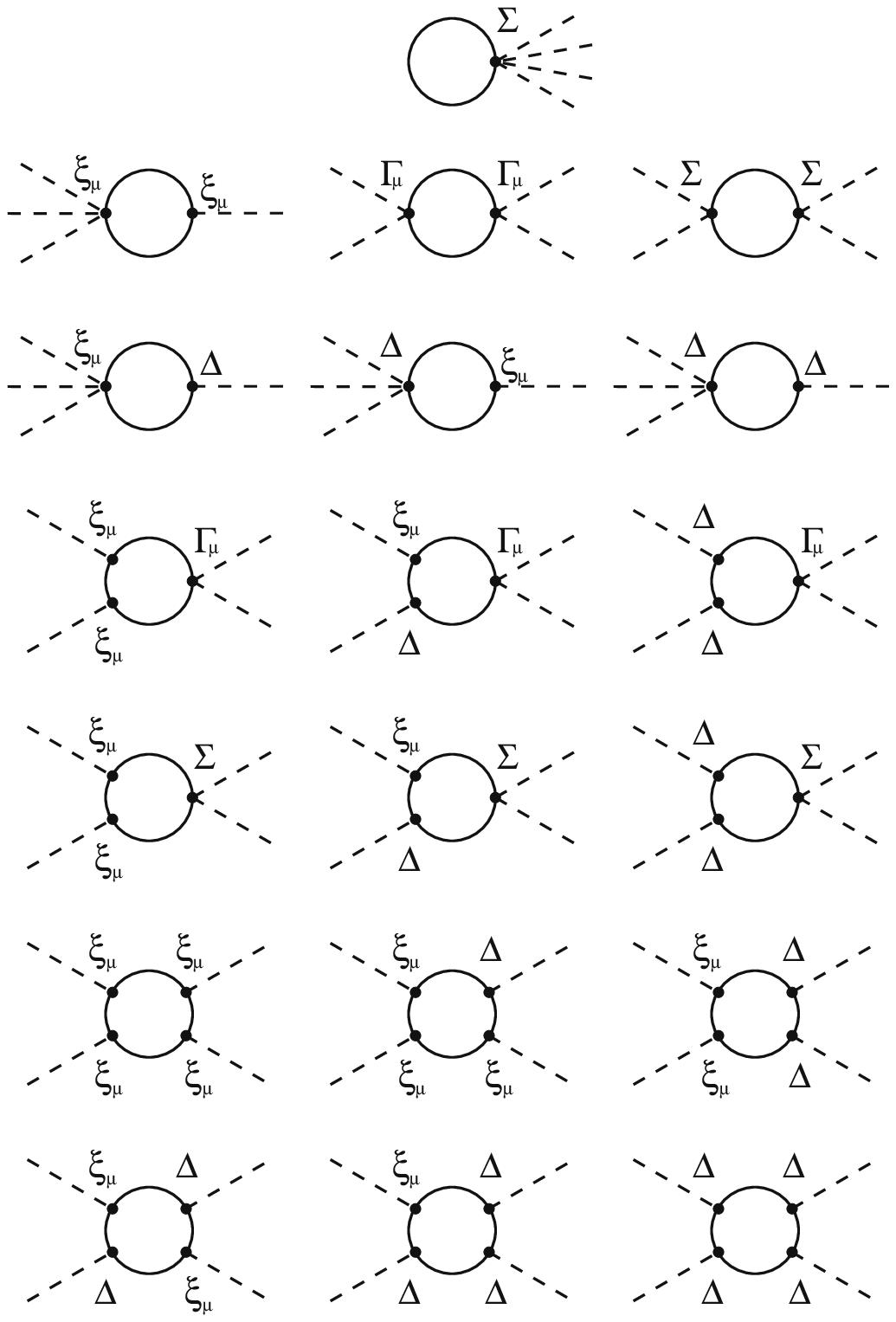}
}
\vskip -1.5cm

\centerline{\Large Figure 5}
\end{figure}


\begin{figure}[bht]
\centerline{
\epsfxsize=5in
\epsfysize=7in
\epsffile{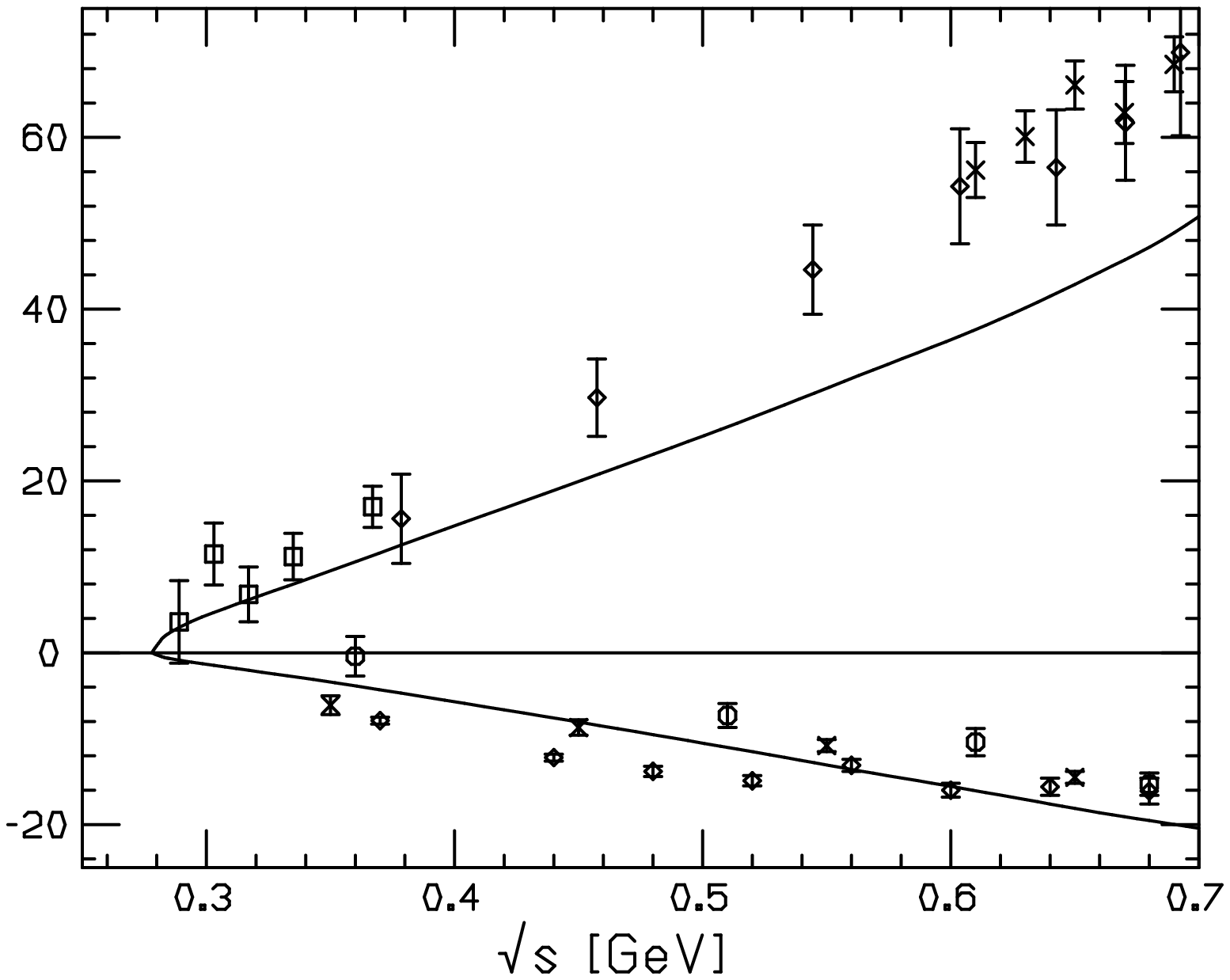}
}
\vskip -1.5cm

\centerline{\Large Figure 6a}
\end{figure}


\begin{figure}[bht]
\centerline{
\epsfxsize=5in
\epsfysize=7in
\epsffile{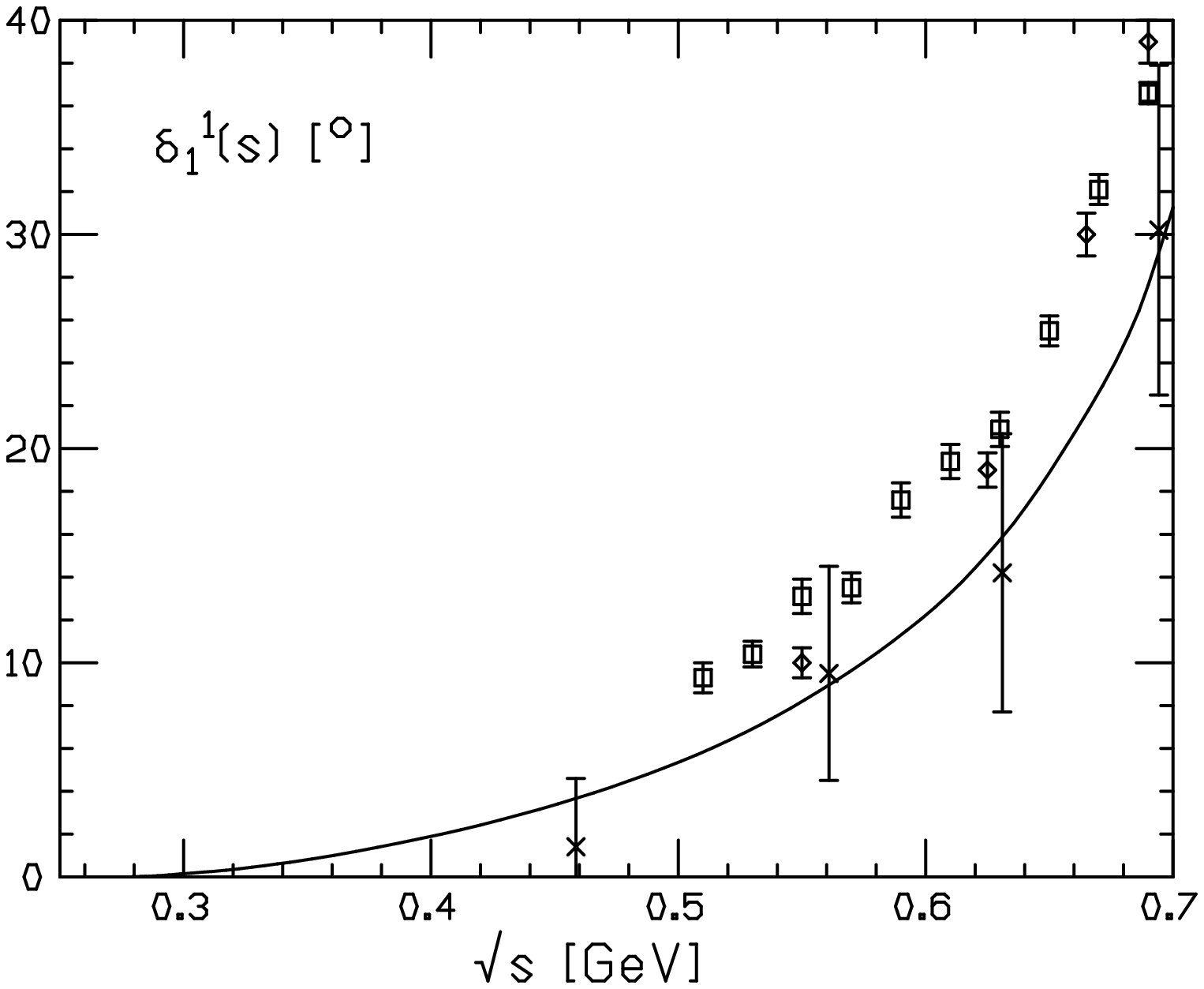}
}

\vskip -1.5cm
\centerline{\Large Figure 6b}
\end{figure}


\begin{figure}[bht]
\centerline{
\epsfxsize=5in
\epsfysize=7in
\epsffile{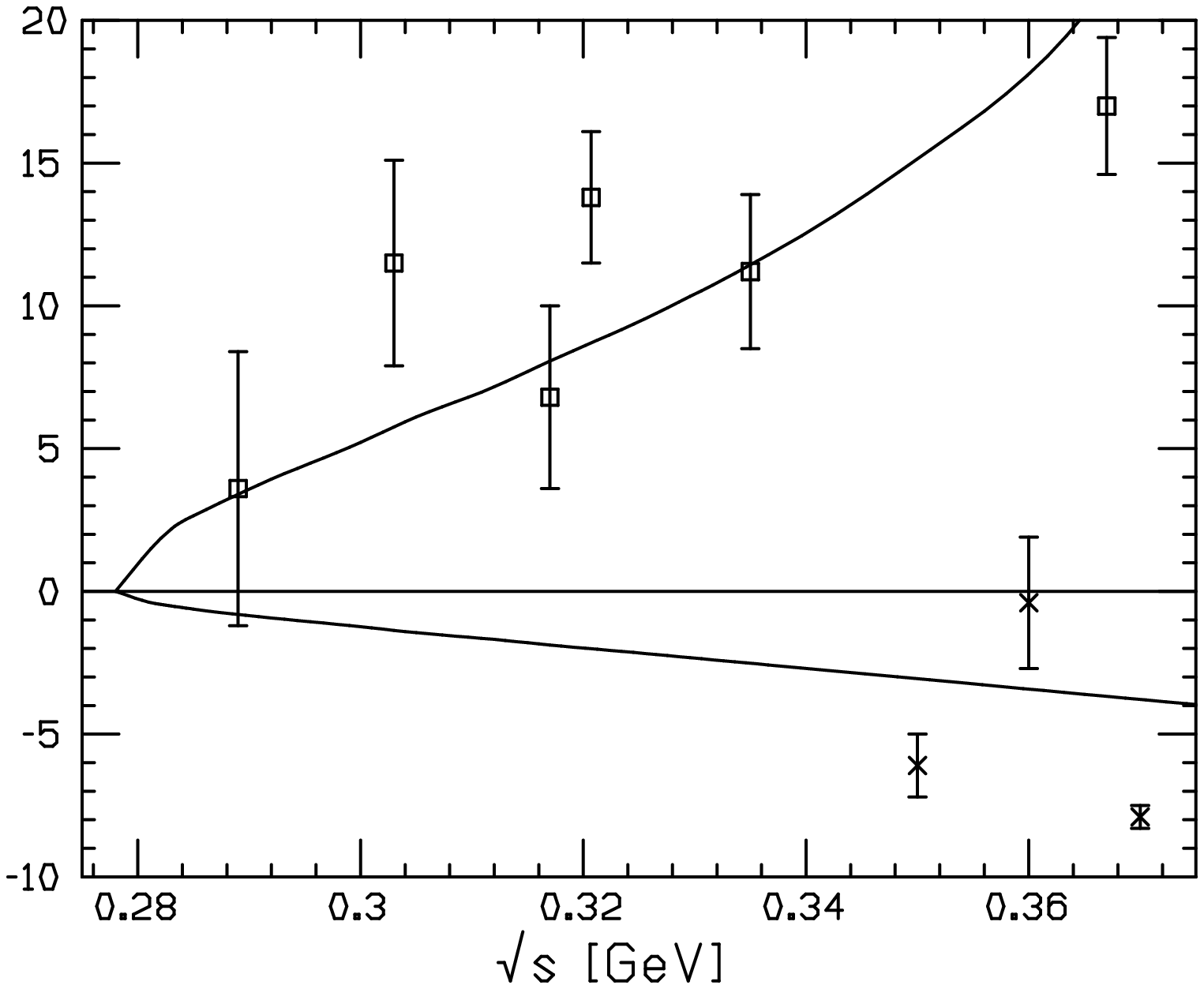}
}

\vskip -1.5cm
\centerline{\Large Figure 6c}
\end{figure}


\begin{figure}[bht]
\centerline{
\epsfxsize=6in
\epsfysize=8in
\epsffile{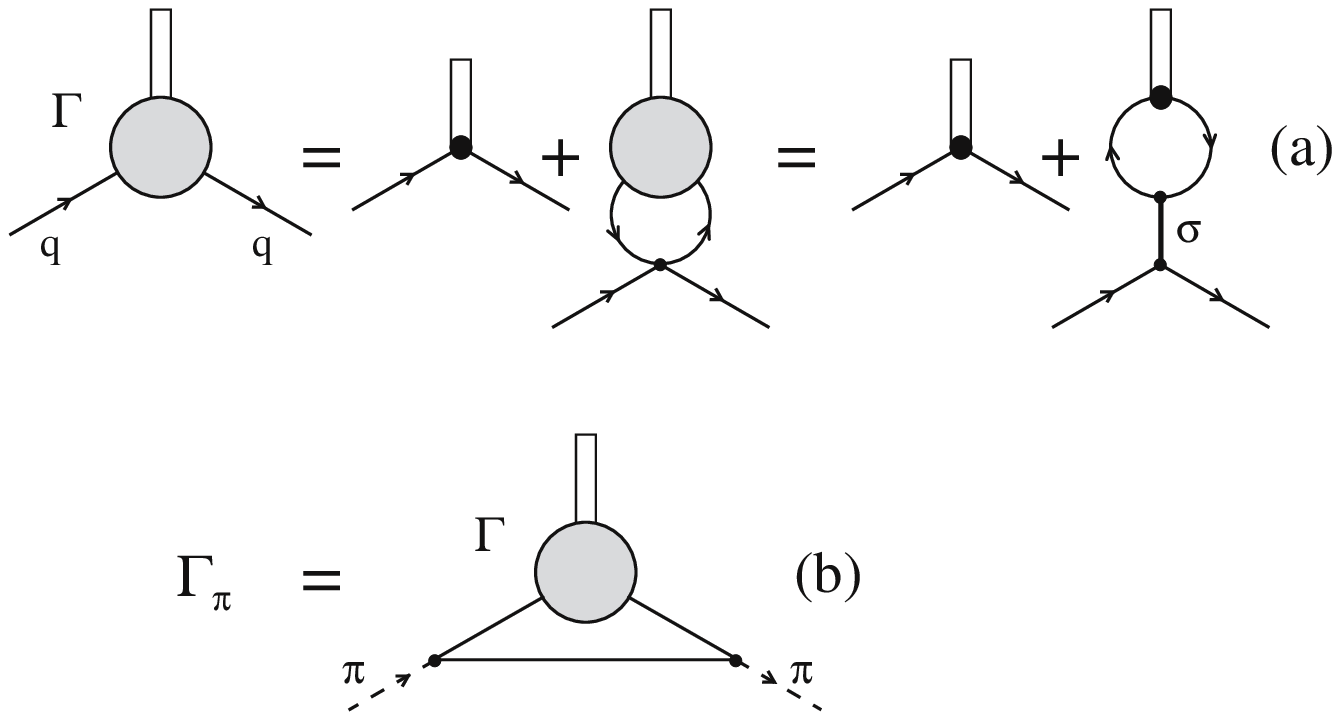}
}
\vskip -1.5cm

\centerline{\Large Figure 7}
\end{figure}



\newpage

\begin{thebibliography}{99}

\bibitem{1}   Y. Nambu and G. Jona-Lasinio, Phys.Rev. 122 (1961) 345;
              124 (1961) 246.
\bibitem{2}   V. Bernard, A. A. Osipov and Ulf-G. Mei\ss ner, Phys.Lett.
              B285 (1992) 119.
\bibitem{3}   V. Bernard, R. L. Jaffe and Ulf-G. Mei\ss ner, Nucl.Phys. B308
              (1988) 753.
\bibitem{4}   V. Bernard and Ulf-G. Mei\ss ner, Nucl.Phys. A489 (1988) 647.
\bibitem{5}   S. Klimt, M. Lutz, U. Vogl and W. Weise, Nucl.Phys. A516 (1990)
              429.
\bibitem{6}   U. Vogl, M. Lutz, S. Klimt and W. Weise, Nucl.Phys. A516 (1990)
              469.
\bibitem{7}   U. Vogl and W. Weise, Progress in Particle and Nuclear Physics
              Vol.26 (1990) 1.
\bibitem{8}   T. Eguchi, Phys.Rev. D14 (1976) 2755; \\
              K. Kikkawa, Progr.Theor.Phys. 56 (1976) 947.
\bibitem{9}   D. Ebert and M. K. Volkov, Yad.Fiz.36 (1982) 1265;
              Z.Phys. C16 (1983) 205.
\bibitem{10}  M. K. Volkov, Ann.Phys. 157 (1984) 282.
\bibitem{11}  A. Dhar and S. R. Wadia, Phys.Rev.Lett. 52 (1984) 959;\\
              A. Dhar, R. Shankar and S. R. Wadia, Phys.Rev. D31 (1985) 3256.
\bibitem{12}  D. Ebert and H. Reinhardt, Nucl.Phys. B271 (1986) 188;
              Phys.Lett. B173 (1986) 453.
\bibitem{13}  D. Espriu, E. de Rafael and J. Taron, Nucl.Phys. B345 (1990) 22.
\bibitem{14}  J. Bijnens, C. Bruno and E. de Rafael, Nucl.Phys. B390 (1993)
              501; \\ for a review, see J. Bijnens, "Chiral Lagrangians and
              Nambu--Jona-Lasinio like models", NORDITA preprint, 1995,
              [hep-ph/9502335].
\bibitem{15}  V. Bernard, Ulf-G.Mei\ss ner, A. H. Blin and B. Hiller,
              Phys.Lett. B253 (1991) 443.
\bibitem{16}  S. Weinberg, Phys.Rev.Lett. 18 (1967) 507.
\bibitem{17}  V. Bernard, A. A. Osipov and Ulf-G.Mei\ss ner, Phys.Lett.
              B324 (1994) 201.
\bibitem{18}  W. Pauli and F. Villars, Rev.Mod.Phys. 21 (1949) 434.
\bibitem{19}  V. Bernard and D. Vautherin, Phys.Rev. D40 (1989) 1615.
\bibitem{20}  C. Sch\"{u}ren, E. Ruiz Arriola and K. Goeke, Nucl.Phys. A547
              (1992) 612.
\bibitem{wein} S. Weinberg, Phys.Rev. 177 (1968) 1568.
\bibitem{cwz} S. Coleman, J. Wess and B. Zumino, Phys.Rev. 177 (1969) 2239.
\bibitem{21}  J. Gasser and H. Leutwyler, Phys.Lett. B125 (1983) 321, 325;
              Ann. of Phys. (NY) 158 (1984) 142.
\bibitem{22}  M. Praszalowicz and G. Valencia, Nucl.Phys. B341 (1990) 27.
\bibitem{23}  R. Haag, Phys.Rev. 112 (1958) 669;\\
              D. Ruelle, Helv.Phys.Acta 35 (1962) 34; \\
              H. I. Borchers, Nuovo cimento 25 (1960) 270.
\bibitem{24}  I. S. R. Chisholm, Nucl.Phys. 26 (1961) 469.
\bibitem{fred} M. Takizawa, K. Kubodera and F. Myhrer, Phys.Lett. B261
               (1991) 221.
\bibitem{25}  C. D. Froggatt and J. L. Petersen, Nucl.Phys. B129 (1977) 89.
\bibitem{27}  J. F. Donoghue, J. Gasser and H. Leutwyler, Nucl.Phys.
              B343 (1990) 341;\\
              J. Gasser and Ulf-G. Mei\ss ner, Nucl.Phys. B357 (1991) 90.
\bibitem{loops} A. Blotz and K. Goeke, Int. J. Mod. Phys. A9 (1994)
  2067; \\ R. Lemmer and R. Tegen, ``Meson Cloud Corrections to the
  Pion Electromagnetic Form Factor in the Nambu Jona-Lasinio Model'',
  J\"ulich preprint 1995.
\bibitem{dat00} L. Rosselet et al., Phys.Rev. D15 (1977) 574; \\
                B. Hyams et al., Nucl. Phys. B63 (1973) 134; \\
                E. Alekseeva et al., Sov.Phys.JETP 55 (1982) 591.
\bibitem{dat02} W. Hoogland et al., Nucl.Phys. B126 (1977) 109; \\
                M.J. Losty et al., Nucl.Phys. B69 (1974) 185; \\
                J.P. Prokup et al., Phys.Rev. D10 (1974) 2055.
\bibitem{dat11} S.D. Protopopescu et al. Phys.Rev. D7 (1973) 1279; \\
                P. Estabrooks and A.D. Martin, Nucl.Phys. B79 (1974) 301; \\
                E. Alekseeva et al., JETP Lett. 29 (1979) 100.
\bibitem{26}  V. Bernard and Ulf-G. Mei\ss ner , Phys.Lett. B266 (1991) 403.
\bibitem{28}  K. L. Au, D. Morgan and M. R. Pennington, Phys.Rev. D35 (1987)
              1633.
\bibitem{29}  S. Weinberg, Phys.Rev.Lett. 17 (1966) 616;\\
              R. W. Griffith, Phys.Rev. 176 (1968) 1705.

\end{thebibliography}
\end{document}